\documentclass[preprint,letter,numberedappendix]{emulateapj}
\usepackage{apjfonts}
\usepackage{graphicx}

\newcommand{\oi}{[O\,{\sc i}]}
\newcommand{\oii}{[O\,{\sc ii}]}
\newcommand{\oiii}{[O\,{\sc iii}]}
\newcommand{\oiv}{[O\,{\sc iv}]}

\newcommand{\Ni}{[N\,{\sc i}]}
\newcommand{\nii}{[N\,{\sc ii}]}

\newcommand{\sii}{[S\,{\sc ii}]}
\newcommand{\siii}{[S\,{\sc iii}]}
\newcommand{\siv}{[S\,{\sc iv}]}

\newcommand{\hei}{He\,{\sc i}}
\newcommand{\heii}{He\,{\sc ii}}

\newcommand{\neii}{[Ne\,{\sc ii}]}
\newcommand{\neiii}{[Ne\,{\sc iii}]}

\newcommand{\ariv}{[Ar\,{\sc iv}]}

\newcommand{\cliii}{[Cl\,{\sc iii}]}

\newcommand{\ha}{H$\alpha$}
\newcommand{\hb}{H$\beta$}

\usepackage{color} 
\newcommand{\kms}{km s$^{-1}$}

\newcommand{\te}{$T_{\epsilon}$}
\newcommand{\Ne}{$n_{\epsilon}$}

\newcommand{\rmxaa}{RMxAA}
\newcommand{\pasa}{PASA}

\received{}
\revised{}
\accepted{}

\slugcomment{}
\shorttitle{Chemical abundances in the halo PN H4-1}
\shortauthors{Otsuka \& Tajistu}

\begin{document}

\title{Chemical abundances in the extremely Carbon and Xenon-rich halo planetary nebula H4-1}

\author{Masaaki Otsuka\altaffilmark{1} and Akito Tajitsu\altaffilmark{2}}
\affil{
$^{1}$Institute of Astronomy and Astrophysics, Academia Sinica
P.O. Box 23-141, Taipei 10617, Taiwan, Republic of China;
otsuka@asiaa.sinica.edu.tw}

\affil{$^{2}$Subaru Telescope, NAOJ, 650 North A'ohoku Place, Hilo,
HI 96720, U.S.A.; tajitsu@subaru.naoj.org}

\begin{abstract}
We performed detailed chemical abundance analysis of the extremely
 metal-poor ([Ar/H]$\sim$--2) halo planetary nebula
 H4-1 based on the multi-wavelength spectra from Subaru/HDS, $GALEX$,
 $SDSS$, and $Spitzer$/IRS and determined the abundances of 10 elements. 
The C and O abundances were derived from collisionally excited lines
 (CELs) and are almost consistent with abundances from recombination lines
 (RLs). We demonstrated that the large discrepancy in the C abundance
 between CEL and RL in H4-1 can be solved using the temperature
 fluctuation model. We reported the first detection of the
 [Xe~{\sc iii}]$\lambda$5846 {\AA} line in H4-1 and determination of its
 elemental abundance ([Xe/H]$>$+0.48). H4-1 is the most Xe-rich PN 
among the Xe-detected PNe. The observed abundances are close to 
the theoretical prediction by a $\sim$2.0 $M_{\odot}$ single star model with
 initially $r$-process element rich ([$r$/Fe]=+2.0 dex). 
The observed Xe abundance would be a product of the $r$-process 
in primordial SNe. The [C/O]-[Ba/(Eu or Xe)] diagram suggests that 
the progenitor of H4-1 shares the evolution
 with two types of carbon-enhanced metal-poor stars (CEMP), CEMP-$r$/$s$ and CEMP-$no$ stars. The progenitor of H4-1
 is a presumably binary formed in an $r$-process rich environment.  
\end{abstract}

\section{Introduction}
Currently, more than 1000 objects are considered planetary nebulae (PNe) 
in the Galaxy, about 14 of which have been identified as halo 
members because of their current location and kinematics
\citep[e.g.,][]{2008ApJ...682L.105O}. 
Halo PNe are interesting objects because they provide direct 
insight into the final evolution of old, low-mass metal-poor halo stars.

The halo PN H4-1 is extremely metal-poor ($Z$$\sim$10$^{-4}$, [Fe/H]=--2.3) 
and C-rich \citep{Torres-Peimbert:1979aa,Kwitter:2003aa}, similar to
 the C-rich halo PNe BoBn1 in the Sagittarius dwarf spheroidal galaxy
 \citep{Otsuka:2010aa,2008ApJ...682L.105O} and K648 in M15 \citep{2002A&A...381.1007R}. The chemical abundances and metallicity 
 of these three PNe imply that their progenitors formed early, 
perhaps $\sim$10-13 Gyrs ago. However, these extremely metal-poor halo 
PNe have an unresolved issue: how did their progenitors evolve
into C-rich PNe?

The initial mass of these three PNe is generally thought to equal $\sim$0.8
$M_{\odot}$, which corresponds to the turn-off stellar mass of 
M15. According to recent stellar evolution models
\citep[e.g.,][]{Lugaro:2012aa,2000ApJ...529L..25F}, 
there are two mechanisms
for stars with $Z$$\lesssim$10$^{-4}$ to become C-rich: 
(1) helium-flash driven deep mixing (He-FDDM) for $<$2.5-3 $M_{\odot}$
stars and (2) third dredge-up (TDU) for $\gtrsim$0.9-1.5 $M_{\odot}$ stars. 
Mechanism (1) is unlikely for these PNe because this mechanism occurs 
in stars with $Z$$\lesssim$6$\times$10$^{-5}$ 
([Fe/H]$\lesssim$--2.5). Therefore, TDU is essential for these PNe 
to become C-rich, implying that it is possible that these halo PNe
evolved from $\gtrsim$0.8-0.9 $M_{\odot}$ single stars. 
We should note that the lower limit mass required for TDU depends largely
on models. The minimum mass for the occurrence of the TDU is
thought to be 1.2-1.5 $M_{\odot}$ \citep[e.g.,][]{Lattanzio:1987aa,Boothroyd:1988aa}.

Even if the progenitor of H4-1 is a $\gtrsim$0.8-0.9 $M_{\odot}$ single star and it
experienced TDU during AGB phase, current low-mass stellar evolution models 
are unlikely to explain the evolution of H4-1; 
the effective temperature of the
central star of H4-1 is between 93\,400 \citep{1997ARep...41..760M} and
132\,000 K \citep{Henry:1996aa}, and the age of the PN is 8400 yrs, assuming a
distance of 25.3 kpc from us \citep{1997ARep...41..760M}. The He-shell burning model
for the initially 0.89 $M_{\odot}$ and $Z$=0.004 stars by
\citet{1994ApJS...92..125V} predicted that it takes $>$50\,000 yrs for a star to
reach 10$^{5}$ K. There is obviously a large discrepancy in evolutional
time scale between the observations and theoretical models.

To explain the evolution of H4-1, we need to consider 
additional mechanisms that shorten the evolutionary timescale, e.g.,  
binary mass-transfer via Roche lobe overflow, 
which was proposed for BoBn1 by \citet{Otsuka:2010aa}. Evidence suggests that binary evolution 
is the most plausible scenario for H4-1. In particular,
\citet{Tajitsu:2004aa} showed that H4-1 has a bright equatorial disk, a bipolar nebula,
and multiple arcs in its molecular hydrogen image. The discovery of these
structures implies that H4-1 would have evolved from a 
binary, similar to carbon-enhanced metal-poor (CEMP) stars 
found in the Galactic halo and these CEMP stars show [Fe/H]$<$--2 \citep[see][]{Beers:2005aa}. \citet{2008AIPC.1016..427O} found that the
location of H4-1 on [C/(Fe or Ar)] versus [(Fe or Ar)/H] diagrams is
in the region occupied by CEMP stars, indicating a similar origin and
evolution.

Elements with atomic number $Z$$>$30 are synthesized in the neutron ($n$)
capturing process in both PN progenitors and supernovae (SNe). In PN 
progenitors, the $n$-flux is much lower, so the $n$-capturing process 
is a slow process ($s$-process). In SNe, because of a much higher 
$n$-flux, the $n$-capturing process is a rapid process ($r$-process). 
Several types of CEMP stars show the enhancements of $s$- and/or
$r$-process elements. Therefore, if we can detect any 
$n$-capture elements, we would be able to obtain important insights 
into the evolution of H4-1 and the chemical environment where its
progenitor was born by comparing the carbon and $n$-capture elements of
H4-1 and those of CEMP stars.

The elements C and O are clearly important with respect to the evolution
of H4-1. However, there is a large discrepancy in the C abundances between 
collisionally excited lines (CELs) and recombination
lines (RLs). \citet{Torres-Peimbert:1979aa} estimated the C number
density using RLs and found that log~$n$(C)/$n$(H)+12=9.39, whereas
\citet{Henry:1996aa} found a value of 8.68 using CELs. Discrepancies in C and O abundances
have already been found in many Galactic PNe \citep[e.g.,][]{Liu:2006aa}. It is
necessary to solve the C discrepancy in H4-1 (the O discrepancy, too if exist).

In this paper, we report a detail chemical abundance analysis of 
H4-1 based on deep 
high-dispersion optical spectra from Subaru/HDS, and archived UV, optical, and
mid-IR spectra from $GALEX$, the Sloan Digital Sky Survey (SDSS), and $Spitzer$/IRS, respectively. In Section 2, 
we describe the observations. In Section 3, we provide the ionic and 
elemental abundances of H4-1 derived from CELs and RLs. The first detection of the
$n$-capture element xenon (Xe) is reported in this section. In Section 4.1, we
discuss the discrepancy in C and O abundances. In Sections 4.2 and 4.3,
we discuss the Xe abundance by comparing with the theoretical nucleosynthesis 
model and the origin and evolution of H4-1 from the view point of
chemical abundances. A summary is given in Section 5.

\section{Observations \& Data Reduction}

\subsection{HDS observations}

\begin{figure}
\includegraphics[width=\columnwidth]{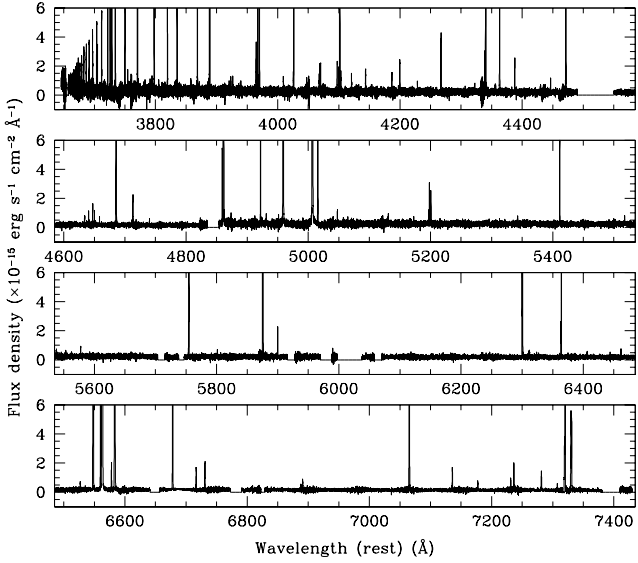}
\caption{The de-reddened HDS spectrum of H4-1. The wavelength is shifted to 
the rest wavelength in air.\label{hds}}
\end{figure}

\begin{deluxetable}{@{}clcccrr@{}}
\tablecolumns{7}
\centering
\tabletypesize{\footnotesize}
\tablecaption{Detected lines and identifications of H4-1 in HDS spectra.\label{hdstab}}
\tablewidth{\columnwidth}
\tablehead{
\colhead{$\lambda_{\rm obs}$}&
\colhead{Ion}&
\colhead{$\lambda_{\rm lab}$}&
\colhead{Comp.}&
\colhead{$f$($\lambda$)}&
\colhead{$I$($\lambda$)}&
\colhead{$\delta$$I$($\lambda$)}\\
\colhead{({\AA})}&
\colhead{}&
\colhead{({\AA})}&
\colhead{}&
\colhead{}&
\colhead{}&
\colhead{}
}
\startdata
3667.25 & H25 & 3669.46 & 1 & 0.334 & 0.355 & 0.047 \\ 
3669.30 & H24 & 3671.48 & 1 & 0.333 & 0.386 & 0.058 \\ 
3671.53 & H23 & 3673.74 & 1 & 0.333 & 0.439 & 0.096 \\ 
3672.35 & He\,{\sc ii} & 3674.84 & 1 & 0.333 & 0.177 & 0.054 \\ 
3674.11 & H22 & 3676.36 & 1 & 0.332 & 0.584 & 0.066 \\ 
3677.16 & H21 & 3679.35 & 1 & 0.332 & 0.564 & 0.061 \\ 
3680.61 & H20 & 3682.81 & 1 & 0.331 & 0.672 & 0.052 \\ 
3684.64 & H19 & 3686.83 & 1 & 0.330 & 0.906 & 0.069 \\ 
3689.34 & H18 & 3691.55 & 1 & 0.329 & 0.911 & 0.062 \\ 
3694.92 & H17 & 3697.15 & 1 & 0.328 & 1.175 & 0.075 \\ 
3701.62 & H16 & 3703.65 & 1 & 0.327 & 1.316 & 0.073 \\ 
3702.81 & He\,{\sc i} & 3704.98 & 1 & 0.327 & 0.647 & 0.047 \\ 
3704.95 & O\,{\sc iii} & 3707.25 & 1 & 0.326 & 0.056 & 0.038 \\ 
3709.73 & H15 & 3711.97 & 1 & 0.325 & 1.513 & 0.075 \\ 
3712.80 & O\,{\sc iii} & 3715.08 & 1 & 0.325 & 0.085 & 0.028 \\ 
3719.69 & H14 & 3721.94 & 1 & 0.323 & 1.971 & 0.090 \\ 
3723.63 & [O\,{\sc ii}] & 3726.03 & 1 & 0.322 & 50.749 & 2.492 \\ 
3723.94 & [O\,{\sc ii}] & 3726.03 & 2 & 0.322 & 40.319 & 2.203 \\ 
 &  &  & Tot. &  & 91.068 & 3.326 \\ 
3726.33 & [O\,{\sc ii}] & 3728.81 & 1 & 0.322 & 42.426 & 1.851 \\ 
3726.69 & [O\,{\sc ii}] & 3728.81 & 2 & 0.322 & 42.795 & 1.841 \\ 
 &  &  & Tot. &  & 85.221 & 2.610 \\ 
3730.62 & He\,{\sc i} & 3732.86 & 1 & 0.321 & 0.103 & 0.031 \\ 
3732.12 & H13 & 3734.37 & 1 & 0.321 & 2.465 & 0.109 \\ 
3747.89 & H12 & 3750.15 & 1 & 0.317 & 2.991 & 0.125 \\ 
3752.40 & O\,{\sc iii} & 3754.70 & 1 & 0.316 & 0.182 & 0.020 \\ 
3768.37 & H11 & 3770.63 & 1 & 0.313 & 3.775 & 0.155 \\ 
3795.61 & H10 & 3797.90 & 1 & 0.307 & 4.840 & 0.194 \\ 
3817.34 & He\,{\sc i} & 3819.60 & 1 & 0.302 & 1.238 & 0.054 \\ 
3831.40 & He\,{\sc i} & 3833.55 & 1 & 0.299 & 0.128 & 0.030 \\ 
3833.08 & H9 & 3835.38 & 1 & 0.299 & 7.301 & 0.286 \\ 
3855.71 & He\,{\sc ii} & 3858.07 & 1 & 0.294 & 0.105 & 0.024 \\ 
3865.16 & He\,{\sc i} & 3867.47 & 1 & 0.291 & 0.136 & 0.021 \\ 
3866.44 & [Ne\,{\sc iii}] & 3868.77 & 1 & 0.291 & 5.323 & 0.225 \\ 
3869.43 & He\,{\sc i} & 3871.83 & 1 & 0.290 & 0.091 & 0.020 \\ 
3886.73 & H8 & 3889.05 & 1 & 0.286 & 9.712 & 0.387 \\ 
3918.27 & C\,{\sc ii} & 3920.68 & 1 & 0.279 & 0.050 & 0.018 \\ 
3921.09 & He\,{\sc ii} & 3923.48 & 1 & 0.278 & 0.124 & 0.016 \\ 
3924.18 & He\,{\sc i} & 3926.54 & 1 & 0.277 & 0.111 & 0.014 \\ 
3962.36 & He\,{\sc i} & 3964.73 & 1 & 0.267 & 0.658 & 0.036 \\ 
3965.08 & [Ne\,{\sc iii}] & 3967.46 & 1 & 0.267 & 1.562 & 0.059 \\ 
3966.01 & He\,{\sc ii} & 3968.43 & 1 & 0.266 & 0.170 & 0.020 \\ 
3967.68 & H7 & 3970.07 & 1 & 0.266 & 14.235 & 0.506 \\ 
4006.90 & He\,{\sc i} & 4009.26 & 1 & 0.256 & 0.191 & 0.021 \\ 
4023.81 & He\,{\sc i} & 4026.18 & 1 & 0.251 & 1.969 & 0.068 \\ 
4065.47 & C\,{\sc iii} & 4067.87 & 1 & 0.239 & 0.162 & 0.015 \\ 
4066.40 & [S\,{\sc ii}] & 4068.60 & 1 & 0.239 & 0.277 & 0.019 \\ 
4067.77 & O\,{\sc ii} & 4070.14 & 1 & 0.239 & 0.261 & 0.018 \\ 
4089.64 & O\,{\sc ii} & 4092.93 & 1 & 0.232 & 0.043 & 0.019 \\ 
4094.83 & N\,{\sc iii} & 4097.35 & 1 & 0.231 & 0.283 & 0.016 \\ 
4097.54 & He\,{\sc ii} & 4100.04 & 1 & 0.230 & 0.249 & 0.039 \\ 
4099.27 & H6 & 4101.73 & 1 & 0.230 & 25.395 & 0.781 \\ 
4100.90 & N\,{\sc iii} & 4103.39 & 1 & 0.229 & 0.231 & 0.028 \\ 
4118.37 & He\,{\sc i} & 4120.81 & 1 & 0.224 & 0.269 & 0.026 \\ 
4126.21 & [Fe\,{\sc ii}] & 4128.75 & 1 & 0.222 & 0.077 & 0.024 \\ 
4141.30 & He\,{\sc i} & 4143.76 & 1 & 0.217 & 0.320 & 0.020 \\ 
4197.27 & He\,{\sc ii} & 4199.83 & 1 & 0.200 & 0.378 & 0.014 \\ 
4264.60 & C\,{\sc ii} & 4267.15 & 1 & 0.180 & 1.118 & 0.035 \\ 
4336.02 & He\,{\sc ii} & 4338.67 & 1 & 0.157 & 0.393 & 0.070 \\ 
4337.85 & H5 & 4340.46 & 1 & 0.157 & 46.910 & 1.069 \\ 
4346.71 & O\,{\sc ii} & 4349.43 & 1 & 0.154 & 0.028 & 0.006 \\ 
4360.52 & [O\,{\sc iii}] & 4363.21 & 1 & 0.149 & 3.677 & 0.167 \\ 
4360.58 & [O\,{\sc iii}] & 4363.21 & 2 & 0.149 & 6.078 & 0.159 \\ 
 &  &  & Tot. &  & 9.755 & 0.231 \\ 
4385.32 & He\,{\sc i} & 4387.93 & 1 & 0.142 & 0.532 & 0.016 \\ 
4387.89 & C\,{\sc iii} & 4390.50 & 1 & 0.141 & 0.071 & 0.020 \\ 
4434.94 & He\,{\sc i} & 4437.55 & 1 & 0.126 & 0.083 & 0.017 \\ 
4468.83 & He\,{\sc i} & 4471.47 & 1 & 0.115 & 4.487 & 0.093 \\ 
4558.82 & Mg\,{\sc i}] & 4562.60 & 1 & 0.087 & 0.031 & 0.007 \\
4631.33 & N\,{\sc iii} & 4634.12 & 1 & 0.065 & 0.083 & 0.011 \\ 
4635.99 & O\,{\sc ii} & 4638.86 & 1 & 0.064 & 0.057 & 0.006 \\ 
4637.80 & N\,{\sc iii} & 4640.64 & 1 & 0.063 & 0.148 & 0.006 \\
4639.03 & O\,{\sc ii} & 4641.81 & 1 & 0.063 & 0.047 & 0.005 \\ 
4644.57 & C\,{\sc iii} & 4647.42 & 1 & 0.061 & 0.232 & 0.007 
\enddata
\end{deluxetable}
\begin{deluxetable}{@{}clcccrr@{}}
\centering
\setcounter{table}{1}
\tablecolumns{7}
\tablecaption{Continued.}
\tablewidth{\columnwidth}
\tablehead{
\colhead{$\lambda_{\rm obs}$}&
\colhead{Ion}&
\colhead{$\lambda_{\rm lab}$}&
\colhead{Comp.}&
\colhead{$f$($\lambda$)}&
\colhead{$I$($\lambda$)}&
\colhead{$\delta$$I$($\lambda$)}\\
\colhead{({\AA})}&
\colhead{}&
\colhead{({\AA})}&
\colhead{}&
\colhead{}&
\colhead{}&
\colhead{}
}
\startdata
4647.39 & C\,{\sc iii} & 4650.25 & 1 & 0.060 & 0.147 & 0.006 \\
4648.66 & O\,{\sc ii} & 4651.33 & 1 & 0.060 & 0.024 & 0.005 \\ 
4655.58 & C\,{\sc iv} & 4658.20 & 1 & 0.058 & 0.065 & 0.004 \\ 
4658.89 & O\,{\sc ii} & 4661.63 & 1 & 0.057 & 0.035 & 0.005 \\ 
4682.60 & He\,{\sc ii} & 4685.68 & 1 & 0.050 & 4.491 & 0.109 \\ 
4682.91 & He\,{\sc ii} & 4685.68 & 2 & 0.050 & 13.819 & 0.199 \\ 
 &  &  & Tot. &  & 18.310 & 0.227 \\ 
4708.49 & [Ar\,{\sc iv}] & 4711.37 & 1 & 0.042 & 0.065 & 0.007 \\ 
4710.36 & He\,{\sc i} & 4713.17 & 1 & 0.042 & 0.554 & 0.013 \\ 
4737.32 & [Ar\,{\sc iv}] & 4740.17 & 1 & 0.034 & 0.059 & 0.004 \\ 
4856.37 & He\,{\sc ii} & 4859.32 & 1 & 0.001 & 0.873 & 0.017 \\ 
4858.37 & H4 & 4861.33 & 1 & 0.000 & 42.807 & 1.844 \\ 
4858.43 & H4 & 4861.33 & 2 & 0.000 & 57.193 & 1.731 \\ 
 &  &  & Tot. &  & 100.000 & 2.529 \\ 
4919.00 & He\,{\sc i} & 4921.93 & 1 & --0.016 & 1.006 & 0.017 \\ 
4928.29 & [O\,{\sc iii}] & 4931.80 & 1 & --0.019 & 0.073 & 0.013 \\ 
4955.89 & [O\,{\sc iii}] & 4958.91 & 1 & --0.026 & 112.614 & 2.513 \\ 
4955.96 & [O\,{\sc iii}] & 4958.91 & 2 & --0.026 & 111.463 & 1.811 \\ 
 &  &  & Tot. &  & 224.077 & 3.097 \\ 
5003.67 & [O\,{\sc iii}] & 5006.84 & 1 & --0.038 & 76.548 & 18.101 \\ 
5003.85 & [O\,{\sc iii}] & 5006.84 & 2 & --0.038 & 578.884 & 17.282 \\ 
5004.51 & [O\,{\sc iii}] & 5006.84 & 3 & --0.038 & 12.015 & 7.893 \\ 
 &  &  & Tot. &  & 667.447 & 26.242 \\ 
5012.69 & He\,{\sc i} & 5015.68 & 1 & --0.040 & 1.907 & 0.035 \\ 
5028.99 & C\,{\sc ii} & 5032.07 & 1 & --0.044 & 0.044 & 0.008 \\ 
5032.83 & [Fe\,{\sc ii}] & 5035.48 & 1 & --0.045 & 0.030 & 0.005 \\ 
5044.74 & He\,{\sc i} & 5047.74 & 1 & --0.048 & 0.153 & 0.007 \\ 
5124.28 & Fe\,{\sc iii} & 5127.39 & 1 & --0.066 & 0.018 & 0.004 \\ 
5127.90 & C\,{\sc iii} & 5130.86 & 1 & --0.067 & 0.079 & 0.004 \\ 
5137.51 & C\,{\sc ii} & 5140.79 & 1 & --0.069 & 0.046 & 0.009 \\ 
5194.45 & [N\,{\sc i}] & 5197.900 & 1 & --0.082 & 0.144 & 0.009 \\ 
5195.10 & [N\,{\sc i}] & 5197.90 & 2 & --0.082 & 0.366 & 0.010 \\ 
 &  &  & Tot. &  & 0.510 & 0.013 \\ 
5196.81 & [N\,{\sc i}] & 5200.26 & 1 & --0.083 & 0.117 & 0.008 \\ 
5197.45 & [N\,{\sc i}] & 5200.26 & 2 & --0.083 & 0.315 & 0.012 \\ 
 &  &  & Tot. &  & 0.432 & 0.015 \\ 
5339.16 & C\,{\sc ii} & 5342.19 & 1 & --0.112 & 0.078 & 0.008 \\ 
5339.19 & C\,{\sc ii} & 5342.43 & 1 & --0.112 & 0.082 & 0.014 \\ 
5408.25 & He\,{\sc ii} & 5411.52 & 1 & --0.126 & 1.119 & 0.024 \\ 
5514.37 & [Cl\,{\sc iii}] & 5517.66 & 1 & --0.145 & 0.062 & 0.007 \\ 
5534.55 & [Cl\,{\sc iii}] & 5537.60 & 1 & --0.149 & 0.062 & 0.007 \\ 
5573.87 & [O\,{\sc i}] & 5577.34 & 1 & --0.156 & 0.046 & 0.014 \\ 
5574.30 & [O\,{\sc i}] & 5577.34 & 2 & --0.156 & 0.075 & 0.011 \\ 
 &  &  & Tot. &  & 0.120 & 0.018 \\ 
5751.02 & [N\,{\sc ii}] & 5754.64 & 1 & --0.185 & 0.588 & 0.031 \\ 
5751.37 & [N\,{\sc ii}] & 5754.64 & 2 & --0.185 & 0.903 & 0.030 \\ 
 &  &  & Tot. &  & 1.491 & 0.043 \\ 
5797.75 & C\,{\sc iv} & 5801.34 & 1 & --0.192 & 0.060 & 0.007 \\ 
5843.18 & [Xe\,{\sc iii}]& 5846.67 & 1 & --0.199 & 0.047 & 0.008 \\ 
        & + He\,{\sc ii}\\
5872.14 & He\,{\sc i} & 5875.62 & 1 & --0.203 & 14.918 & 0.363 \\ 
6070.48 & He\,{\sc ii} & 6074.30 & 1 & --0.232 & 0.033 & 0.022 \\ 
6147.72 & C\,{\sc ii} & 6151.27 & 1 & --0.242 & 0.059 & 0.009 \\ 
6166.94 & He\,{\sc ii} & 6170.69 & 1 & --0.245 & 0.056 & 0.010 \\ 
6230.09 & He\,{\sc ii} & 6233.82 & 1 & --0.254 & 0.023 & 0.009 \\ 
6296.11 & [O\,{\sc i}] & 6300.30 & 1 & --0.263 & 1.198 & 0.038 \\ 
6296.64 & [O\,{\sc i}] & 6300.30 & 2 & --0.263 & 0.520 & 0.113 \\ 
6296.87 & [O\,{\sc i}] & 6300.30 & 3 & --0.263 & 4.681 & 0.177 \\ 
 &  &  & Tot. &  & 6.399 & 0.214 \\ 
6307.03 & He\,{\sc ii} & 6310.85 & 1 & --0.264 & 0.050 & 0.010 \\ 
6308.17 & [S\,{\sc iii}] & 6312.10 & 1 & --0.264 & 0.106 & 0.013 \\ 
6359.54 & [O\,{\sc i}] & 6363.78 & 1 & --0.271 & 0.377 & 0.016 \\ 
6359.98 & [O\,{\sc i}] & 6363.78 & 2 & --0.271 & 0.134 & 0.041 \\ 
6360.30 & [O\,{\sc i}] & 6363.78 & 3 & --0.271 & 1.602 & 0.064 \\ 
 &  &  & Tot. &  & 2.114 & 0.077 \\ 
6402.47 & He\,{\sc ii} & 6406.38 & 1 & --0.277 & 0.079 & 0.008\\
6457.98 & N\,{\sc ii} & 6461.71 & 1 & --0.284 & 0.134 & 0.011  \\
6523.13 & He\,{\sc ii} & 6527.10 & 1 & --0.293 & 0.091 & 0.011 \\ 
6543.91 & [N\,{\sc ii}] & 6548.04 & 1 & --0.296 & 8.636 & 0.348 \\ 
6544.38 & [N\,{\sc ii}] & 6548.04 & 2 & --0.296 & 19.071 & 0.623 \\ 
 &  &  & Tot. &  & 27.707 & 0.714 \\ 
6556.05 & He\,{\sc ii} & 6560.10 & 1 & --0.297 & 0.710 & 0.083 \\ 
6556.16 & He\,{\sc ii} & 6560.10 & 2 & --0.297 & 1.142 & 0.068 \\
 &  &  & Tot. &  & 1.852 & 0.107 
\enddata
\end{deluxetable}
\begin{deluxetable}{@{}clcccrr@{}}
\tablecolumns{7}
\setcounter{table}{1}
\tablecaption{Continued.}
\tablewidth{\columnwidth}
\tablehead{
\colhead{$\lambda_{\rm obs}$}&
\colhead{Ion}&
\colhead{$\lambda_{\rm lab}$}&
\colhead{Comp.}&
\colhead{$f$($\lambda$)}&
\colhead{$I$($\lambda$)}&
\colhead{$\delta$$I$($\lambda$)}\\
\colhead{({\AA})}&
\colhead{}&
\colhead{({\AA})}&
\colhead{}&
\colhead{}&
\colhead{}&
\colhead{}
}
\startdata
6558.82 & H3 & 6562.77 & 1 & --0.298 & 96.472 & 4.710 \\ 
6558.87 & H3 & 6562.77 & 2 & --0.298 & 188.528 & 6.350 \\ 
 &  &  & Tot. &  & 285.000 & 7.906 \\ 
6574.09 & C\,{\sc ii} & 6578.05 & 1 & --0.300 & 0.376 & 0.015 \\ 
6579.25 & [N\,{\sc ii}] & 6583.46 & 1 & --0.300 & 25.484 & 0.965 \\ 
6579.73 & [N\,{\sc ii}] & 6583.46 & 2 & --0.300 & 55.214 & 1.775 \\ 
 &  &  & Tot. &  & 80.697 & 2.021 \\ 
6674.13 & He\,{\sc i} & 6678.15 & 1 & --0.313 & 1.805 & 0.070 \\ 
6674.18 & He\,{\sc i} & 6678.15 & 2 & --0.313 & 1.596 & 0.056 \\ 
 &  &  & Tot. &  & 3.401 & 0.090 \\ 
6712.27 & [S\,{\sc ii}] & 6716.44 & 1 & --0.318 & 0.270 & 0.021 \\ 
6712.70 & [S\,{\sc ii}] & 6716.44 & 2 & --0.318 & 0.235 & 0.018 \\ 
 &  &  & Tot. &  & 0.505 & 0.028 \\ 
6726.50 & [S\,{\sc ii}] & 6730.81 & 1 & --0.320 & 0.170 & 0.014 \\ 
6727.02 & [S\,{\sc ii}] & 6730.81 & 2 & --0.320 & 0.408 & 0.018 \\ 
 &  &  & Tot. &  & 0.577 & 0.023 \\ 
6886.73 & He\,{\sc ii} & 6890.90 & 1 & --0.341 & 0.132 & 0.008 \\ 
7061.04 & He\,{\sc i} & 7065.18 & 1 & --0.364 & 4.345 & 0.160 \\ 
7071.72 & [Xe\,{\sc v}]? & 7076.80 & 1 & --0.366 & 0.020 & 0.007 \\ 
7104.53 & [K\,{\sc iv}] & 7108.90 & 1 & --0.370 & 0.015 & 0.009 \\ 
7111.35 & C\,{\sc ii} & 7115.60 & 1 & --0.371 & 0.020 & 0.007 \\ 
7131.50 & [Ar\,{\sc iii}] & 7135.80 & 1 & --0.374 & 0.348 & 0.016 \\ 
7156.36 & He\,{\sc i} & 7160.61 & 1 & --0.377 & 0.050 & 0.012 \\ 
7173.17 & He\,{\sc ii} & 7177.52 & 1 & --0.379 & 0.112 & 0.007 \\ 
7226.94 & C\,{\sc ii} & 7231.34 & 1 & --0.387 & 0.154 & 0.015 \\ 
7232.09 & C\,{\sc ii} & 7236.42 & 1 & --0.387 & 0.345 & 0.017 \\ 
7277.06 & He\,{\sc i} & 7281.35 & 1 & --0.393 & 0.535 & 0.062 \\ 
7285.33 & [Rb\,{\sc v}]? & 7289.81 & 1 & --0.394 & 0.021 & 0.005 \\ 
7293.77 & He\,{\sc i} & 7298.04 & 1 & --0.395 & 0.020 & 0.006 \\ 
7314.87 & [O\,{\sc ii}] & 7319.25 & 1 & --0.398 & 0.933 & 0.058 \\ 
7315.51 & [O\,{\sc ii}] & 7319.89 & 1 & --0.398 & 0.772 & 0.085 \\ 
7315.99 & [O\,{\sc ii}] & 7320.37 & 2 & --0.398 & 2.193 & 0.108 \\ 
 &  &  & Tot. &  & 2.965 & 0.137 \\ 
7325.26 & [O\,{\sc ii}] & 7329.65 & 1 & --0.400 & 0.753 & 0.064 \\ 
7325.55 & [O\,{\sc ii}] & 7329.94 & 2 & --0.400 & 0.819 & 0.050 \\ 
 &  &  & Tot. &  & 1.572 & 0.082 \\ 
7326.40 & [O\,{\sc ii}] & 7330.79 & 1 & --0.400 & 0.703 & 0.053 \\ 
7326.62 & [O\,{\sc ii}] & 7331.01 & 2 & --0.400 & 0.648 & 0.038 \\ 
 &  &  & Tot. &  & 1.351 & 0.065 \\ 
7342.94 & [V \,{\sc ii}] & 7347.74 & 1 & --0.402 & 0.057 & 0.010 \\ 
7361.03 & O\,{\sc iii} & 7365.35 & 1 & --0.404 & 0.041 & 0.009 \\ 
7369.17 & [V \,{\sc ii}] & 7373.32 & 1 & --0.406 & 0.056 & 0.008 \\ 
7373.23 & [Ni\,{\sc ii}] & 7377.83 & 1 & --0.406 & 0.026 & 0.011 
\enddata
\end{deluxetable}

The spectra of H4-1 were taken using the High-Dispersion Spectrograph
\citep[HDS;][]{2002PASJ...54..855N} attached to one of the two
Nasmyth foci of the 8.2-m Subaru telescope (Program ID: S09A-163S, PI: M.Otsuka).

The red spectra (4600-7500 {\AA}) were obtained in May 2009, when weather conditions were
unfavorable; there were scattered cirrus clouds in the sky and a full moon.
The seeing was $\sim$1.5$''$ from the guider CCD.
An atmospheric dispersion corrector (ADC) was used to minimize the differential
atmospheric dispersion through the broad wavelength region. We set the
slit width to equal $1.\hspace{-2pt}''5$ and chose a 2$\times$2 on-chip binning.
We set the slit length to $7''$, which fit the nebula
and allowed a direct subtraction of sky background from the
object frames. The CCD sampling pitch along the slit length projected on
the sky equaled $\sim0.\hspace{-2pt}''276$ per a binned pixel.
The resolving power reached approximately $R$ $>$33\,000, which is derived from
the mean full width at half maximum (FWHM) of narrow Th-Ar
comparisons and night sky lines. We took a series of 1800 s exposures and the
total exposure time was 12\,600 s (7 exposure frames). 
In addition, we took a series of four 300 s exposures to measure the fluxes of strong lines such as
{\oiii}$\lambda$5007. For the flux calibration, blaze function 
correction, and airmass correction, we observed the standard star Hz44
three times at different airmasses.

The blue spectra (3600-5400 {\AA}) were obtained in July 2009 using the same settings employed
in the May 2009 observations, except for the slit width
($1.\hspace{-2pt}''2$). 
The seeing during these observations was 0.62$''$. We took four 1800 s exposure frames on 4 July and a single 600 s frame on 9 July.

Data reduction and emission line analysis were performed mainly with
the long-slit reduction package noao.twodspec in IRAF\footnote[3]{IRAF
is distributed by the National Optical Astronomy Observatories, operated by the Association of Universities for Research in
Astronomy (AURA), Inc., under a cooperative agreement with the National
Science Foundation.}. The resulting signal-to-noise (S/N) ratios at the peak of the detected emission lines in the spectra 
were $>$5 after
subtracting the sky background, whereas that at the continuum was $>$40
at the blue spectrum and $>$50 at the red spectrum before subtracting the sky, even at the edges of each echelle order.

The line were de-reddened using the
formula: $I$($\lambda$) = $F$($\lambda$)$\times$10$^{c({\rm
H}\beta)f(\lambda)}$, where $I$($\lambda$) is the de-reddened line
flux, $F$($\lambda$) is the observed line flux, 
$f$($\lambda$) is the interstellar extinction parameter at $\lambda$
computed by the reddening law of \citet{1989ApJ...345..245C} with
$R_{V}$=3.1, and $c$(H$\beta$) is the reddening coefficient at H$\beta$,
respectively. The values of $F$({\hb})
were 1.85$\times$10$^{-13}$ $\pm$ 8.49$\times$10$^{-15}$ in blue and 
2.13$\times$10$^{-13}$ $\pm$ 4.60$\times$10$^{-15}$ erg s$^{-1}$
cm$^{-2}$ in red. Hereafter, X(--Y) means
X$\times$10$^{\rm -Y}$. We measured $c$({\hb}) by comparing 
the observed Balmer line ratios of H$\gamma$ (blue spectrum)
or {\ha} (red spectrum) to {\hb} with the theoretical ratio of 
\citet{1995MNRAS.272...41S}, assuming the electron temperature {\te}=10$^{4}$
K and the electron density {\Ne}=10$^{4}$ cm$^{-3}$ in the Case B assumption. 
The value of $c$({\hb}) was 0.07$\pm$0.05 for the blue spectra and 0.11$\pm$0.04 for the red
spectra.

Flux scaling was performed using all of the emission lines detected in
the overlap region between the blue and the red spectra. The de-reddened
fluxes relative to $I$({\hb})=100 in both spectra are
coincident within 11$\%$ of each other. The combined de-reddened spectrum
is presented in Fig. \ref{hds}. The observed wavelength at the time of 
observation was corrected to the averaged line-of-sight heliocentric
radial velocity of --181.35$\pm$0.35 {\kms} (root-mean-square of
the residuals: 4.00 {\kms}) among over 120 lines detected in the HDS spectrum.

The detected lines are listed in Table \ref{hdstab}. To ensure accuracy in the
identification, we checked the presence of all of
the detected lines and removed ghost emissions in the two-dimensional
spectra. When measuring
fluxes of the emission lines, we assumed that the line profiles were all 
Gaussian and we applied multiple Gaussian fitting techniques. We listed
the observed wavelength and de-reddened relative fluxes 
of each Gaussian component (indicated by Comp.ID number in the fourth
column of Table \ref{hdstab}) with respect to the
de-reddened H$\beta$ flux of 100.
The line-profiles of most the detected lines can be fit by a
single Gaussian component. For the lines composed of multiple
components (e.g., [O\,{\sc ii}]$\lambda$3726.03 {\AA}), we list the de-reddened relative fluxes of each component
and list the sum of these components (indicated by Tot.) in the last
line.

\subsection{The total {\hb} flux}
To normalize the line fluxes relative to $F$({\hb}) in the 
$GALEX$ and $Spitzer$/IRS spectra, the $F$(\hb) of the whole nebula is
needed. 

A measurement of $F$({\hb})=3.16(--13) erg s$^{-1}$
cm$^{-2}$ was taken for the entire nebula by \citet{Kwitter:2003aa}
using a slit width of 5{\arcsec}. They also determined $c$({\hb})=0.10. 

The SDSS spectrum for H4-1 (SDSS; Object ID:
631018077386964992) provides another measurement, taken using a 3{\arcsec} diameter optical fiber. 
A total {\hb} flux of 4.57(--13)$\pm$3.13(--15) erg s$^{-1}$ 
cm$^{-2}$ was measured by applying
an aperture correction factor of 2.07, which is listed in the SDSS
webpage and is from the $r$-band flux ratio between the image 
and the fiber spectra. A value of $c$({\hb})=0.10$\pm$0.04 was derived by comparing
the observed $F$({\hb})/$F$(H$\gamma$) ratio to the theoretical value
for the case of {\te}=10$^{4}$ K and {\Ne}=10$^{4}$ cm$^{-3}$, because the
H$\alpha$ line was saturated.

We used a {\hb} flux value of 3.86(--13) erg s$^{-1}$ 
cm$^{-2}$, which is the average between the values from the SDSS and 
\citet{Kwitter:2003aa}.

\subsection{{\it GALEX} archive data}

\begin{deluxetable}{@{}lcccc@{}}
\tablecolumns{5}
\centering
\tablecaption{The detected lines in the $GALEX$ spectra.\label{galex}}
\tablewidth{\columnwidth}
\tablehead{
\colhead{$\lambda_{\rm lab}$}    &
\colhead{Ion} &
\colhead{$f$($\lambda$)}&
\colhead{$F$($\lambda$)$^{a}$}&
\colhead{$I$($\lambda$)}\\
\colhead{(\AA)}&
\colhead{}&
\colhead{}&
\colhead{(erg s$^{-1}$ cm$^{-2}$)}&
\colhead{
[$I$({\hb})=100]
}            
}
\startdata
1549/51	&C~{\sc iv}	  &1.238&1.32(--12)$\pm$5.36(--14)&3.42(+2)$\pm$1.89(+1)\\
1640	&He~{\sc ii}	  &1.177&4.87(--13)$\pm$4.38(--14)&1.26(+2)$\pm$1.23(+1)\\
1660/66	&$[$O~{\sc iii}$]$&1.167&2.10(--13)$\pm$7.83(--14)&5.43(+1)$\pm$2.04(+1)\\
1906/09	&C~{\sc iii}$]$	  &1.257&6.04(--12)$\pm$7.73(--14)&1.56(+3)$\pm$6.18(+1)\\
2326	&$[$C~{\sc ii}$]$ &1.364&1.68(--12)$\pm$1.86(--14)&4.35(+2)$\pm$1.70(+1)\\
2470	&$[$O~{\sc ii}$]$ &1.045&1.94(--14)$\pm$7.00(--15)&5.10(0)$\pm$1.82(0)
\enddata
\tablenotetext{a}{We adopted the $F$({\hb})=3.86(--13) erg s$^{-1}$ cm$^{-2}$.
}
\end{deluxetable}

We analyzed the Galaxy Evolution Explorer
\citep[$GALEX$;][]{Martin:2005aa} 
slit-less prism spectra to estimate the CEL 
C$^{+,2+,3+}$ and O$^{+}$ abundances using the [C\,{\sc ii}] $\lambda$2326 {\AA}, 
C\,{\sc iii}$]$ $\lambda$1906/09 {\AA}, C\,{\sc iv} 
$\lambda$1549/51 {\AA}, and {\oii} $\lambda$2470 {\AA} lines. 
The {\it GALEX} spectra were retrieved from the multi-mission
archive at the STScI (MAST; GALEX Object ID: 2555724756316858403). 
We used both FUV (1344-1786 {\AA}) and NUV (1771-2831 {\AA}) band spectra.
We did not correct the reddening because the observed flux ratio of
He~{\sc ii} $F$(1640 {\AA})/$F$(4686 {\AA}) = 6.88$\pm$0.62 was comparable
to the theoretical ratio of He\,{\sc ii}
$I$($\lambda$1640)/($\lambda$4686)=6.56 in the case of {\te}=10$^{4}$ K and {\Ne}=10$^{4}$
cm$^{-3}$ as given by \citet{1995MNRAS.272...41S}. 

The observed and normalized fluxes of the detected lines are listed in 
Table \ref{galex}.

\subsection{$Spitzer$/IRS archive data}

\begin{deluxetable}{@{}rlccr@{}}
\tablecolumns{5}
\centering
\tablecaption{The detected atomic lines in the $Spitzer$ spectra.\label{spitzer}}
\tablewidth{\columnwidth}
\tablehead{
\colhead{$\lambda_{\rm lab}$}    &
\colhead{Ion} &
\colhead{$f$($\lambda$)}&
\colhead{$F$($\lambda$)$^{a}$}&
\colhead{$I$($\lambda$)}\\
\colhead{($\mu$m)}&
\colhead{}&
\colhead{}&
\colhead{(erg s$^{-1}$ cm$^{-2}$)}&
\colhead{
[$I$({\hb})=100]
}            
}
\startdata
10.51 & {\siv}      & --0.960 & 4.97(--15)$\pm$4.06(--16) & 1.21$\pm$0.15 \\ 
12.40 & H\,{\sc i}         & --0.980 & 4.33(--15)$\pm$3.94(--16) & 1.04$\pm$0.13 \\ 
12.81 & {\neii}     & --0.983 & 1.62(--15)$\pm$1.85(--16) & 0.39$\pm$0.06 \\ 
15.56 & {\neiii}    & --0.985 & 1.48(--14)$\pm$3.89(--16) & 3.56$\pm$0.34 \\ 
16.12 & {\hei}        & --0.984 & 3.08(--15)$\pm$2.36(--16) & 0.74$\pm$0.09 \\ 
17.63 & {\hei}       & --0.981 & 6.12(--15)$\pm$6.05(--16) & 1.47$\pm$0.20 \\ 
18.71 & {\siii}     & --0.981 & 3.80(--15)$\pm$2.34(--16) & 0.91$\pm$0.10 \\ 
25.89 & {\oiv}      & --0.989 & 1.29(--13)$\pm$8.57(--16) & 30.87$\pm$2.84 
\enddata
\tablenotetext{a}{
We adopted the $F$({\hb})=3.86(--13) erg s$^{-1}$ cm$^{-2}$.
}
\end{deluxetable}

The spectra were taken on June 12, 2007 using the Infrared 
Spectrograph \citep[$IRS$;][]{Houck:2004aa} with the SL (5.2-14.5 $\mu$m), 
SH (9.9-19.6 $\mu$m), and LH (18.7-37.2 $\mu$m) modules 
(AOR Keys: 18628864 and 186291202; PI: J. Bernard-Salas). 
We used the reduction package SMART v.8.2.5 provided by the IRS 
team at Cornell University \citep{Higdon:2004aa} and IRSCLEAN 
provided by the $Spitzer$ Science Center. For SH and LH spectra, we subtracted the sky background
using off-set spectra. We scaled the SL data up to SH\&LH in the
overlapping wavelength region, and then re-scaled the scaled 5.2-37.2
$\mu$m to match the Wide-field Infrared Survey Explorer ($WISE$) bands 3
($\lambda_{c}$=11.56 $\mu$m) and 4 ($\lambda_{c}$=22.09 $\mu$m) average flux densities (1.91(--13) erg s$^{-1}$ cm$^{-2}$ $\mu$m$^{-1}$ in band 3 and 
3.16(--13) erg s$^{-1}$ cm$^{-2}$ $\mu$m$^{-1}$ in band 4).
 
The line fluxes of detected atomic lines are listed in Table
\ref{spitzer}. We derived $c$({\hb})=0.03$\pm$0.04 by comparing the
observed intensity ratio of H~{\sc i} $F$(12.4 $\mu$m)/$F$({\hb}) to
the theoretical value of \citet{1995MNRAS.272...41S} for the case of 
$T_{\epsilon}$=10$^{4}$ K and $n_{\epsilon}$=10$^{4}$ cm$^{-3}$ (1.04). The line at 12.4 $\mu$m is the
complex of H\,{\sc i} $n$=9-7, and $n$=11-8. We used 
the interstellar extinction function given by \citet{Fluks:1994aa}.

\section{Results}

\subsection{CELs Diagnostics}

\begin{figure}
\centering
\includegraphics[width=\columnwidth]{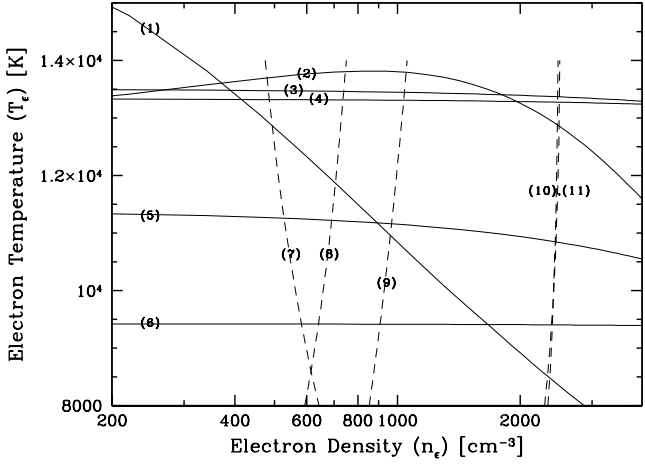}
\caption{$n_{\epsilon}$-$T_{\epsilon}$ diagram. Each curve is labeled with an ID
 number given in Table \ref{diagno_table}.  The solid lines indicate
 diagnostic lines of the $T_{\epsilon}$. The broken lines indicate diagnostic
 lines of the $n_{\epsilon}$. \label{diagno_figure}}
\end{figure}

\begin{deluxetable*}{@{}lclrr@{}}
\tablecolumns{5}
\centering
\tablecaption{Plasma Diagnostics.\label{diagno_table}}
\tablewidth{\textwidth}
\tablehead{
\colhead{Parameters}    &
\colhead{ID} &
\colhead{Diagnostic}&
\colhead{Ratio}&
\colhead{Result}                  
}
\startdata
$T_{\epsilon}$ &(1)&$[$O~{\sc ii}$]$($\lambda$3726+$\lambda$3729)/($\lambda$7320+$\lambda$7330) & 26.459$\pm$1.051$^{a}$ & 12000$\pm$330 \\ 
(K)            &(2)&$[$S~{\sc iii}$]$($\lambda$18.7$\mu$m)/($\lambda$9531) & 0.364$\pm$0.099 & 12\,600$\pm$2200 \\ 
               &(3)&$[$Ne~{\sc iii}$]$($\lambda$15.5$\mu$m)/($\lambda$3869+$\lambda$3967) &0.518$\pm$0.052 &13\,400$\pm$500\\   
               &(4)&$[$O~{\sc iii}$]$($\lambda$4959+$\lambda$5007)/($\lambda$4363) & 91.390$\pm$3.465 & 13\,280$\pm$200 \\ 
               &(5)&$[$N~{\sc ii}$]$($\lambda$6548+$\lambda$6583)/($\lambda$5755) & 72.701$\pm$2.541 & 11\,300$\pm$180 \\ 
              &(6)&$[$O~{\sc i}$]$($\lambda$6300+$\lambda$6363)/($\lambda$5577) & 70.704$\pm$10.527 & 9490$\pm$510 \\ 
\cline{2-5}
              &   &He~{\sc i}($\lambda$7281)/($\lambda$6678) & 0.157$\pm$0.019 & 7430$\pm$2190 \\ 
             &   &(Balmer Jump)/(H 11)                      &0.103$\pm$0.205&11\,970$\pm$2900\\
\hline
$n_{\epsilon}$&(7) &$[$N~{\sc i}$]$($\lambda$5198)/($\lambda$5200) & 1.182$\pm$0.051 & 590$\pm$80 \\ 
(cm$^{-3}$)   &(8) &$[$O~{\sc ii}$]$($\lambda$3726)/($\lambda$3729) & 1.069$\pm$0.051 & 710$\pm$110 \\ 
              &(9) &$[$S~{\sc ii}$]$($\lambda$6716)/($\lambda$6731) & 0.875$\pm$0.059 & 1030$\pm$210 \\ 
              &(10)&$[$Ar~{\sc iv}$]$($\lambda$4711)/($\lambda$4740) & 1.097$\pm$0.143 & 2750$\pm$1460 \\ 
              &(11)&$[$Cl~{\sc iii}$]$($\lambda$5517)/($\lambda$5537) & 1.001$\pm$0.158 & 2840$\pm$1250 \\ 
\cline{2-5}
              &    &Balmer decrement                                  &                 &100-2000
\enddata

\tablenotetext{a}{Corrected recombination contribution for $[$O\,{\sc ii}$]$ $\lambda\lambda$7320/30.}
\end{deluxetable*}

Electron temperatures and densities were derived
from a variety of line diagnostic ratios by solving for level
populations using a multi-level ($>$ 5 levels) atomic model.
The observed diagnostic line ratios are listed in Table \ref{diagno_table}. The numbers in the second column indicate 
the ID of each curve in the {\Ne}-{\te} diagram presented in Fig. \ref{diagno_figure}. The solid lines indicate diagnostic lines 
for the electron temperatures, while the broken lines
are electron density diagnostics. The third, fourth, and last columns in
Table \ref{diagno_table} give 
the diagnostic lines, their line ratios, and the resulting {\Ne} and {\te},
respectively. The value of {\te}({\siii}) was estimated using the observed 
{\siii}$\lambda$18.7 $\mu$m and {\siii}$\lambda$9531 {\AA} lines. We
adopted $I$({\siii}$\lambda$9531)=2.51 measured by \citet{Kwitter:2003aa}.

For the {\oii}$\lambda\lambda$7320/30 line, we
subtracted recombination contamination from O$^{2+}$
using the following equation given by \citet{2000MNRAS.312..585L}:
\begin{equation}
\label{roii}
\frac{I_{R}(\rm [O\,{\sc II}]\lambda\lambda7320/30)}{I(\rm H\beta)} =
9.36\left(\frac{T_{\epsilon}}{10^4}\right)^{0.44}\times\frac{\rm O^{2+}}{\rm H^{+}}.
\end{equation}
\noindent
Using O$^{2+}$ ionic abundances derived from the 
O~{\sc ii} lines and {\te}=10\,000 K, we estimated that  
$I_{R}$({\rm {\oii}}$\lambda\lambda$7320/30)=0.16, which is a negligible recombination contamination. 
Because we could not detect the N\,{\sc ii} and pure O~{\sc iii} 
recombination lines, we did not estimate the contribution 
of N$^{2+}$ to the {\nii}$\lambda$5744 line and of O$^{3+}$ to 
the {\oiii}$\lambda$4363 line, respectively.

First, we estimated {\Ne} in {\te}=12\,000 K for all
density diagnostic lines. The values of {\te}({\oiii}), {\te}({\neiii}), and 
{\te}({\siii}) were calculated using {\Ne}=2790 cm$^{-3}$, which is
the averaged {\Ne} between {\Ne}({\ariv}) and {\Ne}({\cliii}). 
We calculated {\te}({\nii}) and {\te}({\oii}) using
{\Ne}({\oii}), and {\te}({\oi}) using {\Ne}({\Ni}).

Our estimated {\te} and {\Ne} are comparable to those 
provided by \citet{Kwitter:2003aa}, who estimated 
{\te}({\oiii})=12\,300 K and {\te}({\nii})=10\,200 K. However, their {\te}({\oii}) (6100 K) 
was much lower than our calculation, and they also only gave {\Ne}({\sii}) 
(400 cm$^{-3}$).

\subsection{RL Diagnostics}

\begin{figure}
\centering
\includegraphics[width=\columnwidth]{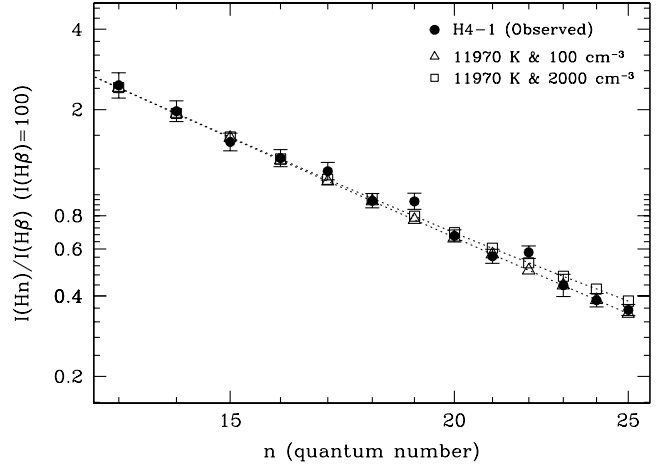}
\caption{
Plot of the intensity ratio of the higher order Balmer lines
to {\hb} (Case B assumption) with the theoretical intensity
 ratios in {\te}=11\,970 K and {\Ne}=100 and 2000 cm$^{-3}$.
\label{bal}}
\end{figure}

We calculated {\te} using the ratio of the Balmer discontinuity to
$I$(H11). The method used to estimate the temperature {\te}(BJ) is
explained in \citet{2001MNRAS.327..141L}.

In the case of {\Ne}=100
cm$^{-3}$, the {\hei} electron temperatures {\te}({\hei}) were derived from the ratio 
He\,{\sc i} $I$($\lambda$7281)/$I$($\lambda$6678) using the emissivity
of {\hei} from \citet{1999ApJ...514..307B}.

The intensity ratio of a high-order Balmer line Hn (where n is
the principal quantum number of the upper level) to a lower-order Balmer line (e.g., {\hb}) is also sensitive to the electron density.
In Fig. \ref{bal}, we plot the ratios of higher-order Balmer lines to
{\hb} with comparisons to theoretical values from
\citet{1995MNRAS.272...41S} 
for values of {\te}(BJ) and {\Ne}=100 and and 2000 cm$^{-3}$, respectively. The electron density 
in the RL emitting region is in the above range. 

The {\te} and {\Ne} derived from the RL diagnostics are also summarized 
in Table \ref{diagno_table}.

\subsection{CEL Ionic Abundances}

\begin{figure}
\centering
\includegraphics[width=\columnwidth]{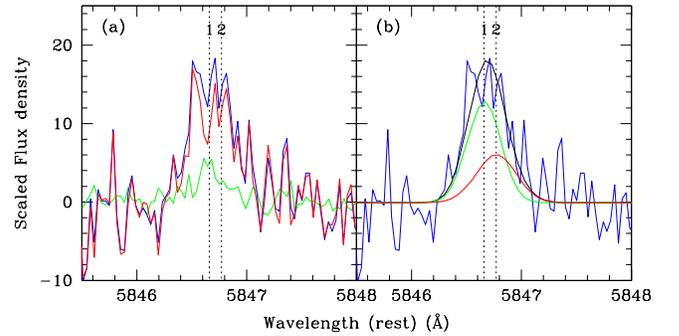}
\caption{(a) The observed line profile at $\lambda$5846.7
 {\AA} (blue line). The dotted lines are the rest wavelength of the He\,{\sc
 ii}$\lambda$5846.7 and [Xe\,{\sc iii}]$\lambda$5846.8 lines, which are
 indicated with the numbers 1 and 2, respectively. The line is composed of He\,{\sc ii}$\lambda$5846.7
 and [Xe\,{\sc iii}]$\lambda$5846.8 transitions. The green line is the expected He\,{\sc
 ii}$\lambda$5846.7 profile made by scaling intensity down and shifting
 wavelength of He\,{\sc ii}$\lambda$6074.3 to He\,{\sc
 ii}$\lambda$5846. The red line is the solo [Xe\,{\sc
 iii}]$\lambda$5846.8 (method I).
(b) The observed line profile at $\lambda$5846.7
 {\AA} (blue line). This line could be represented by 
two Gaussian components; the green line is He\,{\sc ii}$\lambda$5846.7 and
 the red line is [Xe\,{\sc iii}]$\lambda$5846.8. The black line is the
 sum of these two components (method II, See text in detail).
\label{xe3}}
\end{figure}

\begin{deluxetable}{@{}ccl@{}}
\tablecolumns{3}
\centering
\tablecaption{The Adopting electron temperatures and densities.\label{tene}}
\tablewidth{\columnwidth}
\tablehead{
\colhead{{\te} (K)}&
\colhead{{\Ne} (cm$^{-3}$)}&
\colhead{Ions}
}
\startdata
9400    &590  &N$^{0}$, O$^{0}$\\
11\,300 &710  &C$^{+}$, N$^{+}$, O$^{+}$ \\
11\,300 &1030 &S$^{+}$\\
13\,280 &2790 &C$^{2+}$, C$^{3+}$, O$^{2+}$, O$^{3+}$, Ne$^{+}$,
               Ne$^{2+}$\\
        &     &S$^{2+}$, S$^{3+}$, Cl$^{2+}$, Ar$^{2+}$, Ar$^{3+}$, Xe$^{2+}$
\enddata
\end{deluxetable}

\begin{deluxetable}{cccc}
\tablecolumns{4}
\centering
\tablecaption{Ionic Abundances From CELs.\label{celabund}}
\tablewidth{\columnwidth}
\tablehead{
\colhead{X$^{\rm m+}$} &
\colhead{$\lambda_{\rm lab}$}&
\colhead{$I$($\lambda_{\rm lab}$)}    &
\colhead{X$^{\rm m+}$/H$^{+}$} 
}
\startdata         
C$^{+}$ & 2326 {\AA} & 4.35(+2)$\pm$1.70(+1)  & {\bf 4.00(--4)$\pm$3.88(--5)} \\ 
C$^{2+}$ & 1906/9 {\AA} & 1.56(+3)$\pm$6.18(+2) &  {\bf 5.51(--4)$\pm$5.23(--5)} \\ 
C$^{3+}$ & 1549/51 {\AA} & 3.42(+2)$\pm$1.89(+1) &  {\bf 8.85(--5)$\pm$1.04(--5)} \\ 
N$^{0}$ & 5197.9 {\AA} & 5.10(--1)$\pm$1.33(--2) &  8.88(--7)$\pm$1.77(--7) \\ 
  & 5200.3 {\AA} & 4.32(--1)$\pm$1.47(--2)       &  8.93(--7)$\pm$1.72(--7) \\ 
 &  &     & {\bf 8.91(--7)$\pm$1.75(--7)} \\ 
N$^{+}$ & 5754.6 {\AA} & 1.49(0)$\pm$4.30(--2) &  1.17(--5)$\pm$8.85(--7) \\ 
  & 6548.0 {\AA} & 2.77(+1)$\pm$7.14(--1) & 1.19(--5)$\pm$5.37(--7) \\ 
  & 6583.5 {\AA} & 8.07(+1)$\pm$2.02(0) &  1.17(--5)$\pm$5.25(--7) \\ 
 &  &   &  {\bf 1.18(--5)$\pm$5.33(--7)} \\ O$^{0}$ & 5577.3 {\AA} & 1.20(--1)$\pm$1.76(--2) & 1.49(--5)$\pm$4.84(--6) \\ 
  & 6300.3 {\AA} & 6.40(0)$\pm$2.14(--1) & 1.51(--5)$\pm$2.80(--6) \\ 
  & 6363.8 {\AA} & 2.11(0)$\pm$7.72(--2) & 1.56(--5)$\pm$2.91(--6) \\ 
 &  &  & {\bf  1.52(--5)$\pm$2.86(--6)} \\ 
O$^{+}$ & 3726.0 {\AA} & 9.11(+1)$\pm$3.33(0) & 4.55(--5)$\pm$3.14(--6) \\ 
  & 3728.8 {\AA} & 8.52(+1)$\pm$2.61(0) & 4.58(--5)$\pm$3.12(--6) \\ 
  & 7320/30 {\AA} & 6.66(0)$\pm$2.11(--1) &  4.98(--5)$\pm$4.39(--6) \\ 
  & 2470 {\AA} & 5.10(0)$\pm$1.82(0) & 5.07(--5)$\pm$1.82(--5) \\ 
 &  & &  {\bf 4.58(--5)$\pm$3.16(--6)} \\ 
O$^{2+}$ & 1660/66 {\AA} & 5.43(+1)$\pm$2.03(+1) &  2.91(--4)$\pm$1.12(--4) \\ 
  & 4361.2 {\AA} & 9.76(0)$\pm$2.31(--1) &  9.66(--5)$\pm$7.63(--6) \\ 
  & 4931.8 {\AA} & 7.33(--2)$\pm$1.28(--2) &  7.93(--5)$\pm$1.42(--5) \\ 
  & 4958.9 {\AA} & 2.24(+2)$\pm$3.10(0) &  9.46(--5)$\pm$4.01(--6) \\ 
 & 5006.8 {\AA} & 6.67(+2)$\pm$2.62(+1) &  9.76(--5)$\pm$5.50(--6) \\ 
 &  &  &      {\bf 9.69(--5)$\pm$5.15(--6)} \\ 
O$^{3+}$ & 25.9 $\mu$m & 3.09(+1)$\pm$2.84(0) & {\bf 7.56(--6)$\pm$7.01(--7)} \\ 
Ne$^{+}$ & 12.8 $\mu$m & 3.91(--1)$\pm$5.70(--2) &{\bf 4.50(--7)$\pm$6.58(--8)} \\ 
Ne$^{2+}$ & 3868.8 {\AA} & 5.32(0)$\pm$2.25(--1) & 2.26(--6)$\pm$1.43(--7) \\ 
  & 3967.5 {\AA} & 1.56(0)$\pm$5.90(--2) & 1.59(--6)$\pm$9.58(--8) \\ 
  & 15.6 $\mu$m& 3.56(0)$\pm$3.39(--1)  & 2.03(--6)$\pm$1.94(--7) \\ 
 &  &    &{\bf 2.08(--6) $\pm$ 1.53(--7)} \\ 
S$^{+}$ & 4068.6 {\AA} & 2.77(--1)$\pm$1.92(--2) &  5.19(--8)$\pm$4.44(--9) \\ 
  & 6716.4 {\AA} & 5.05(--1)$\pm$2.76(--2) &  2.23(--8)$\pm$1.46(--9) \\ 
  & 6730.8 {\AA} & 5.77(--1)$\pm$2.31(--2) &  2.20(--8)$\pm$1.15(--9) \\ 
 &  &     &{\bf 2.21(--8)$\pm$1.29(--9)} \\ 
S$^{2+}$ & 6312.1 {\AA} & 1.06(--1)$\pm$1.29(--2) & 8.15(--8)$\pm$1.07(--8) \\ 
  & 18.7 $\mu$m& 9.15(--1)$\pm$1.00(--1) & 8.17(--8)$\pm$9.01(--9) \\ 
 &  &  &{\bf 8.17(--8)$\pm$9.19(--9)} \\ 
S$^{3+}$ & 10.5 $\mu$m& 1.77(0)$\pm$2.41(--1) &{\bf 4.77(--8)$\pm$6.50(--9)} \\ 
Cl$^{2+}$ & 5517.7 {\AA} & 6.16(--2)$\pm$7.19(--3) & 3.71(--9)$\pm$4.52(--10) \\ 
 & 5537.6 {\AA} & 6.16(--2) $\pm$ 6.60(--3)  & 3.60(--9) $\pm$ 4.05(--10) \\ 
 &  &   &  {\bf 3.65(--9) $\pm$ 4.29(--10)} \\ 
Ar$^{2+}$ & 7135.8 {\AA} & 3.48(--1)$\pm$1.56(--2) &{\bf 1.75(--8)$\pm$9.27(--10)} \\ 
Ar$^{3+}$ & 4711.4 {\AA} & 6.52(--2)$\pm$7.00(--3) &7.72(--9)$\pm$8.88(--10) \\ 
  & 4740.2 {\AA} & 5.94(--2)$\pm$4.35(--3) &  7.56(--9)$\pm$6.33(--10) \\ 
 &  &   & {\bf  7.65(--9)$\pm$7.67(--10)} \\ 
Xe$^{2+}$ & 5846.7 {\AA} & $>$2.59(--2) & {\bf $>$2.95(--10)}$^{\rm a}$\\
Ba$^{+}$  & 6141.7 {\AA} & $<$9.83(--3) & {\bf $<$3.23(--11)}$^{\rm b}$
\enddata
\tablenotetext{a}{from method I.}
\tablenotetext{b}{The value from 1-$\sigma$ flux density around 6141.7
 {\AA}. See text in detail.}
\end{deluxetable}

We obtained 18 ionic abundances: C$^{+,2+,3+}$, N$^{0,1+}$, 
O$^{0,1+,2+,3+}$, Ne$^{+,2+}$, S$^{+,2+,3+}$, Ar$^{2+,3+}$, 
Cl$^{2+}$, and Xe$^{2+}$. The estimates of 
C$^{+,3+}$, O$^{3+}$, Cl$^{2+}$, and Xe$^{2+}$ are reported for the first time. The ionic abundances were calculated by 
solving the statistical equilibrium equations for more than five levels
in adopting {\te} and {\Ne}. We used electron temperatures and densities 
for each ion that were determined using CEL plasma diagnostics. The
adopted {\te} and {\Ne} values for 
each ion are listed in Table \ref{tene}.

The derived ionic abundances are presented in Table
\ref{celabund}. The last column contains the resulting ionic abundances, 
X$^{\rm m+}$/H$^{+}$, and their probable errors, which include errors from 
line intensities, electron temperature, and electron density. 
In the last line of each ion's line series, we present the adopted ionic abundance and its
error. These values are estimated from the weighted mean of the line intensity.

When calculating the C$^{+}$ abundance, we subtracted contamination 
from [O~{\sc iii}] $\lambda$2321 {\AA} to [C~{\sc ii}] $\lambda$2326 
{\AA} based on the theoretical intensity ratio of {\oiii} 
$I$(2326 {\AA})/$I$(4363 {\AA}) = 0.236. To determine the final S$^{+}$
abundance, we excluded the values derived from the trans-aural line
{\sii}$\lambda$4069 {\AA} because this line would be contaminated from
C~{\sc iii} $\lambda$4068 {\AA}. For O$^{2+}$, we excluded
{\oiii}$\lambda$1660/66 {\AA} because of its large uncertainty.

\citet{2003AJ....125..265D} reported the upper limit intensities of 
[Ne\,{\sc ii}]$\lambda$12.8 $\mu$m and [S\,{\sc iv}]$\lambda$10.5 
$\mu$m and estimated Ne$^{+}$/H$^{+}$$<$4.6(--6) and
S$^{3+}$/H$^{+}$$<$3.8(--7) using the mid-IR spectra taken by
TEXES/IRTF. The $Spitzer$ observations improved their estimates.

This is the first detection of [Xe\,{\sc iii}]$\lambda$5846.8 {\AA} 
($^{3}P_{2}{-}^{1}D_{2}$) in H4-1. 
The detection of this element is very rare because there are only a handful
of detection cases in PNe \citep{2007ApJ...659.1265S,2009PASA...26..339S,Garcia-Rojas:2012aa,Otsuka:2010aa}.

The local continuum subtracted [Xe\,{\sc iii}]$\lambda$5846.8
line profile is presented in Fig. \ref{xe3}. In high-excitation PNe such
as H4-1, this line appears to be blended with
He\,{\sc ii}$\lambda$5846.7 ($n$=31-5; Pfund series). The blue line in
Fig. \ref{xe3}(a) indicates the observed line profile. The $[$Fe\,{\sc ii}$]$
$\lambda$5847.3 line possibly contributes to the emission line 
at $\lambda$5846, however, this line has poorer wavelength agreement 
and our photo-ionization model of H4-1 using {\sc CLOUDY} code
\citep{Ferland:1998aa} predicted that the intensity ratio of $I$([Fe\,{\sc
ii}]$\lambda$5846.7) to $I$(H$\beta$) is $<$1(--8). 
Therefore, the emission line at $\lambda$5846 would be a combination of the 
[Xe\,{\sc iii}] and He~{\sc ii} lines.

We attempted to obtain the solo [Xe\,{\sc iii}]$\lambda$5846.8 line flux by
subtracting the contribution from He\,{\sc ii}$\lambda$5846.7 using
the following method (method I). First, we scaled the intensity of
He\,{\sc ii}$\lambda$6074.3 down to He\,{\sc ii}$\lambda$5846.7 using
the theoretical ratio of $I$($\lambda$5846.7)/$I$(6074.3), which is 0.271 in
the case of {\te}=11\,970 K (={\te}(BJ)) and 1000 cm$^{-3}$ according
to \citet{1995MNRAS.272...41S}. Second, we shifted the intensity peak
wavelength of this scaled line to 5846.7 {\AA}. The predicted He\,{\sc
ii}$\lambda$5846.7 is indicated by the green line in
Fig. \ref{xe3}(a). Next, we subtracted this shifted line from the 5846
{\AA} line. Finally, we obtained an intensity of the [Xe\,{\sc iii}] $\lambda$5846.8 line
of 3.63(--2)$\pm$1.04(--2), as indicated by the red line in
Fig. \ref{xe3}(a).

We also attempted to obtain the [Xe\,{\sc iii}]$\lambda$5846.8 using
Gaussian fitting, as shown in Fig. \ref{xe3}(b) (method II). As
constraints, we fix the rest wavelengths of the intensity peak and FWHM
velocities to equal 5846.7 {\AA} and 18.33 {\kms} for He\,{\sc
ii}$\lambda$5846.7, respectively, and 5846.8 and 21.90 {\kms} for [Xe\,{\sc
iii}]$\lambda$5846.8. Here, 18.33 {\kms} and 21.90 {\kms}
correspond to the FWHM velocities of {\heii}$\lambda$6074.3 and
{\cliii}$\lambda$5537 (I.P.=23.81 eV). Because the I.P. of Xe$^{2+}$
(21.21 eV) is very close to Cl$^{2+}$, we assumed that the FWHM velocity
of $[$Xe\,{\sc iii}$]$ is almost identical to that of
{\cliii}$\lambda$5537. In method II, we obtained $I$($[$Xe\,{\sc
iii}$]$$\lambda$5846.8)$>$1.42(--2).

Based on the above analyses, it is probable that [Xe\,{\sc iii}]
$\lambda$5846.8 would be in the HDS spectrum. However, we need to carefully consider
the estimation of the $[$Xe\,{\sc iii}$]$$\lambda$5846.8
line intensity; because we could not detect any Pfund series He\,{\sc ii} lines
with an upper $n$ from 29 to 21 and $>$32, He\,{\sc ii}$\lambda$5846.7
may be not in the HDS spectrum or it may be present at a very low intensity. The 
emission line at 5846.8 {\AA} would be mostly from the [Xe\,{\sc iii}]
$\lambda$5846.8 line. In this work, we determined the lower limit Xe$^{2+}$ and Xe
abundances using method I based on the transition probabilities of
\citet{Biemont:1995aa} and the collisional impacts of \citet{1998A&AS..128..581S}.

We could not retrieve strong lines for Ba~{\sc ii} $\lambda$4934.6 {\AA} and
6141.7 {\AA} from the HDS spectrum because both lines were deeply affected by moonlight.
Therefore, we calculated an upper limit for the expected Ba$^{+}$ abundance
of 3.23(--10) using the 1-$\sigma$ flux density at Ba\,{\sc
ii}$\lambda$6141.7 ($I$(6141.7 {\AA})/$I$({\hb})=9.83(--3),
$I$({\hb})=100) with {\te}=11\,300 K, {\Ne}=1030 cm$^{-3}$, and FWHM
velocity = 19 {\kms}. In the Ba$^{+}$ calculation, we used transition
probabilities from \citet{2002JPCRD..31..217K} and collisional impacts
from \citet{1998A&AS..128..581S}

\subsection{RL Ionic Abundances}

\begin{deluxetable}{@{}llccc@{}}
\tablecolumns{5}
\tablecaption{Ionic Abundances From RLs.\label{rlabund}}
\tablewidth{\columnwidth}
\tablehead{
\colhead{X$^{\rm m+}$} & 
\colhead{Multi.} & 
\colhead{$\lambda_{\rm lab}$}& 
\colhead{$I$($\lambda_{\rm lab}$)} & 
\colhead{X$^{\rm m+}$/H$^{+}$} 
}
\startdata
He$^{+}$ & V11 & 5875.62 {\AA} & 1.49(+1)$\pm$3.63(--1) & 1.00(--1)$\pm$4.09(--2) \\ 
 & V14 & 4471.47 {\AA} & 4.49(0)$\pm$9.31(--2) & 8.46(--2)$\pm$3.20(--2) \\ 
 & V46 & 6678.15 {\AA} & 3.40(0)$\pm$9.01(--2) &7.97(--2)$\pm$3.27(--2) \\ 
 & V48 & 4921.93 {\AA} & 1.01(0)$\pm$1.70(--2) &7.52(--2)$\pm$3.34(--2) \\ 
 & V51 & 4387.93 {\AA} & 5.32(--2)$\pm$1.64(--2) &8.68(--2)$\pm$3.86(--2) \\ 
 &  &  &   &{\bf 9.30(--2)$\pm$3.77(--2)} \\ 
He$^{2+}$ & 3.4 & 4685.68 {\AA} & 1.83(+1)$\pm$2.27(--1) & {\bf 1.54(--2)$\pm$2.04(--3)} \\ 
C$^{2+}$ & V4 & 3920.68 {\AA} & 4.97(--2)$\pm$1.84(--2) & 1.17(--3)$\pm$4.94(--4) \\ 
 & V6 & 4267.15 {\AA} & 1.12(0)$\pm$3.50(--2) &1.15(--3)$\pm$1.80(--4) \\ 
  & V16.04 & 6151.27 {\AA} & 5.93(--2)$\pm$9.11(--3) & 1.36(--3)$\pm$2.79(--4) \\ 
 & V17.04 & 6461.71 {\AA} & 1.34(--1)$\pm$1.07(--2) & 1.33(--3)$\pm$2.59(--4) \\ 
 & V17.06 & 5342.19 {\AA} & 7.85(--2)$\pm$7.72(--3) & 1.49(--3)$\pm$2.99(--4) \\ 
 & V17.06 & 5342.43 {\AA} & 8.23(--2)$\pm$1.43(--2) & 1.56(--3)$\pm$3.85(--4) \\ 
 &  &  &   &{\bf 1.21(--3)$\pm$2.19(--4)} \\ 
C$^{2+}$ & V1 & 4647.42 {\AA} & 2.32(--1)$\pm$7.15(--3) & 3.64(--4)$\pm$4.37(--5) \\ 
  & V1 & 4650.25 {\AA} & 1.47(--1)$\pm$5.89(--3) & 3.86(--4)$\pm$4.74(--5) \\ 
 & V16 & 4067.87 {\AA} & 1.62(--1)$\pm$1.51(--2) & 4.25(--4)$\pm$7.22(--5) \\ 
 & V16 & 4070.14 {\AA} & 2.61(--1)$\pm$1.76(--2) & 3.82(--4)$\pm$6.00(--5) \\ 
 &  &  &   &{\bf 3.86(--4)$\pm$5.55(--5)} \\ 
C$^{3+}$ & V1 & 4658.20 {\AA} & 6.48(--2)$\pm$4.19(--3) &{\bf  1.55(--5)$\pm$2.45(--6)} \\ 
N$^{3+}$ & V2 & 4634.12 {\AA} & 8.27(--2)$\pm$1.11(--2) & 6.51(--5)$\pm$1.22(--5) \\ 
  & V2 & 4640.64 {\AA} & 1.48(--1)$\pm$5.84(--3) & 6.50(--5)$\pm$8.88(--6) \\ 
 &  &    & &{\bf 6.50(--5)$\pm$1.01(--5)} \\ 
O$^{2+}$ & V1 & 4638.86 {\AA} & 5.71(--2)$\pm$5.99(--3) & 3.51(--4)$\pm$5.74(--5) \\ 
  & V1 & 4641.81 {\AA} & 4.72(--2)$\pm$4.90(--3) & 1.81(--4)$\pm$2.32(--5) \\ 
  & V1 & 4651.33 {\AA} & 2.36(--2)$\pm$5.42(--3) & 1.40(--4)$\pm$3.66(--5) \\ 
  & V1 & 4661.63 {\AA} & 3.53(--2)$\pm$5.14(--3) & 1.96(--4)$\pm$3.77(--5) \\ 
  & V2 & 4349.43 {\AA} & 2.80(--2)$\pm$5.97(--3) & 1.41(--4)$\pm$3.49(--5) \\ 
 &  &  &   &{\bf 1.69(--4)$\pm$3.40(--5)} \\ 
 O$^{3+}$ & V2 & 3754.70 {\AA} &1.82(--1)$\pm$1.97(--2) & 3.47(--4)$\pm$5.80(--5) \\ 
  & V2 & 3757.24 {\AA} & 7.79(--2)$\pm$1.27(--2) &3.35(--4)$\pm$6.92(--5) \\ 
 &  &  &   &{\bf 3.43(--4)$\pm$6.13(--5)} 
\enddata
\end{deluxetable}

The estimated RL ionic abundances are listed in Table \ref{rlabund}. The
 calculations of C$^{3+,4+}$, N$^{3+}$, and O$^{2+,3+}$ abundances were
 performed for the first time. The Case B assumption applies to lines 
from levels that have the same spin as the ground state, and the 
Case A assumption applies to lines of other multiplicities. 
In the last line of each ion's line series, we present the adopted 
ionic abundance and the error estimate from the line intensity-weighted mean. 
Because the RL ionic abundances are insensitive to the electron density 
under $\lesssim$10$^{8}$ cm$^{-3}$, we adopted {\Ne}=10$^{4}$ cm$^{-3}$ 
for He$^{2+}$, C$^{2+,3+,4+}$, N$^{3+}$, and O$^{2+,3+}$ and 
{\Ne}=10$^{2}$ cm$^{-3}$ for He$^{+}$. The emission coefficients are the
 same as those used in \citet{Otsuka:2010aa}.

We detected multiplet V2 N\,{\sc iii} lines, however, these lines
are not recombination lines, but resonance lines. Because we detected the 
O\,{\sc iii} resonance line, the intensity of the resonance line N\,{\sc iii} 
$\lambda$374.36 {\AA} (2$p$ $^{2}P^{0}$-3$d$ $^{2}D$) is enhanced by
O\,{\sc iii} resonance lines at the wavelength of 374.11 {\AA} (2$p^{2}$
$^{3}P$-3$s^{3}$ $P^{0}$). The line intensities of the multiplet V2 lines
can be enhanced by O\,{\sc iii} lines. Furthermore, the detected V2 O\,{\sc iii}
lines might be excited by the Bowen fluorescence mechanism or by the charge 
exchange of O$^{3+}$ and H$^{0}$ instead of by recombination. 
Therefore, we did not use the N$^{3+}$ and O$^{3+}$ abundances to
determine elemental RL N and O abundances.

We detected the multiplet V1 and V2 O\,{\sc ii}
lines. \citet{Ruiz:2003aa}, \citet{Peimbert:2005aa}, and \citet{Garcia-Rojas:2009aa}
pointed out that the upper levels of the transitions in the V1 O\,{\sc
ii} line are not in local thermal equilibrium (LTE) for {\Ne}$<$ 10\,000
cm$^{-3}$, and that the abundances derived from each individual line could
differ by a factor of $\sim$4 \citep{Garcia-Rojas:2009aa}. However, the V2
lines are not affected by non-LTE effects. Because H4-1 is a low-density PN
($<$3000 cm$^{-3}$), we performed the non-LTE corrections
using equations (8) through (10) from \citet{Peimbert:2005aa} with
{\Ne}=2000 cm$^{-3}$. The resulting O$^{2+}$ abundances determined using the V1
and V2 lines are in good agreement, except for the O\,{\sc ii}
$\lambda$4638.86 line. This line was therefore excluded in the final determination of the RL O$^{2+}$ abundance.

\subsection{Elemental Abundances}

\begin{deluxetable}{@{}lccl@{}}
\tablecolumns{4}
\tablecaption{Adopted ionization correction factors (ICFs).\label{icf}}
\tablewidth{\columnwidth}
\tablehead{
\colhead{X}&
\colhead{Line}&
\colhead{ICF(X)}&
\colhead{X/H}
}
\startdata
He &RL &1&He$^{+}$+He$^{2+}$\\
C  &CEL &$\rm \frac{1}{1-\left(\frac{C^{4+}}{C}\right)_{RL}}$&ICF(C)(C$^{+}$+C$^{2+}$+C$^{3+}$)\\
   &RL &$\rm \frac{1}{1-\left(\frac{N^{+}}{N}\right)_{CEL}}$&ICF(C)(C$^{2+}$+C$^{3+}$+C$^{4+}$)\\
N &CEL &$\rm \left(\frac{O}{O^{+}}\right)_{CEL}$&ICF(N)N$^{+}$\\
O  &CEL & 1 &O$^{+}$+O$^{2+}$+O$^{3+}$\\
   &RL &$\rm \left(\frac{O}{O^{2+}}\right)_{CEL}$&ICF(O)O$^{2+}$\\
Ne &CEL &$\rm \left(\frac{O}{O^{+}+O^{2+}}\right)_{CEL}$&ICF(Ne)(Ne$^{+}$+Ne$^{2+}$)\\
S  &CEL &1 &$\rm S^{+}+S^{2+}+S^{3+}$\\
Cl &CEL &$\rm {\left(\frac{Ar}{Ar^{2+}}\right)}$ &ICF(Cl)Cl$^{2+}$\\
Ar &CEL &$\rm \frac{1}{1-\left(\frac{N^{+}}{N}\right)_{CEL}}$&ICF(Ar)$\rm \left(Ar^{2+}+Ar^{3+}\right)$\\
Xe &CEL &$\rm {\left(\frac{Ar}{Ar^{2+}}\right)}$ &ICF(Xe)Xe$^{2+}$\\
Ba &CEL &1                                       &ICF(Ba)Ba$^{+}$
\enddata
\end{deluxetable}

\begin{deluxetable*}{@{}lccrcrc@{}}
\tablecolumns{7}
\centering
\tablecaption{The Elemental Abundances From CEL and RLs.\label{abund}}
\tablewidth{\textwidth}
\tablehead{
\colhead{X} &
\colhead{Types of}  & 
\colhead{X/H} & 
\colhead{log(X/H)+12} & 
\colhead{[X/H]} & 
\colhead{log(X$_{\odot}$/H)+12} & 
\colhead{ICF(X)}   \\ 
\colhead{}  &
\colhead{Emissions} 
}
\startdata
He & RL & 1.08(--2)$\pm$3.78(--2) & 11.04$\pm$0.15 & +0.11$\pm$0.15 & 10.93$\pm$0.01 & 1.00 \\ 
C & RL & 2.33(--3)$\pm$6.91(--4) & 9.37$\pm$0.13 & +0.98$\pm$0.13 & 8.39$\pm$0.04 & 1.44$\pm$0.38 \\ 
C & CEL & 1.04(--3)$\pm$4.42(--4) & 9.02$\pm$0.18 & +0.63$\pm$0.19 & 8.39$\pm$0.04 & 1.01$\pm$0.42 \\ 
N & CEL & 3.85(--5)$\pm$3.56(--6) & 7.59$\pm$0.04 & --0.24$\pm$0.06 & 7.83$\pm$0.05 & 3.28$\pm$0.26 \\ 
O & RL & 2.63(--4)$\pm$7.54(--5) & 8.42$\pm$0.12 & --0.27$\pm$0.13 & 8.69$\pm$0.05 & 1.55$\pm$0.32 \\ 
O & CEL & 1.50(--4)$\pm$6.09(--6) & 8.18$\pm$0.02 & --0.51$\pm$0.05 & 8.69$\pm$0.05 & 1.00 \\ 
Ne & CEL & 2.67(--6)$\pm$5.87(--7) & 6.43$\pm$0.10 & --1.44$\pm$0.14 & 7.87$\pm$0.10 & 1.05$\pm$0.22 \\ 
S & CEL & 1.36(--7)$\pm$1.01(--8) & 5.13$\pm$0.03 & --2.06$\pm$0.05 & 7.19$\pm$0.04 & 1.00 \\ 
Cl & CEL & 7.57(--9)$\pm$2.24(--9) & 3.88$\pm$0.13 & --1.62$\pm$0.33 & 5.50$\pm$0.30 & 2.07$\pm$0.56 \\ 
Ar & CEL & 3.63(--8)$\pm$9.65(--9) & 4.56$\pm$0.12 & --1.99$\pm$0.14 & 6.55$\pm$0.08 & 1.45$\pm$0.38 \\ 
Xe & CEL & $>$5.05(--10)           & $>$2.75 & $>$+0.48 & 2.27$\pm$0.02 & 1.94$\pm$0.51\\
Ba & CEL & $<$3.23(--10)           & $<$2.51 & $<$+0.33 &2.18$\pm$0.03  & 1
\enddata
\end{deluxetable*}

\begin{deluxetable*}{@{}lccccccccccc@{}}
\tablecolumns{11}
\centering
\tabletypesize{\footnotesize}
\tablecaption{Elemental Abundances derived by previous works and by this work. \label{past}}
\tablewidth{\textwidth}
\tablehead{
\colhead{References}& 
\colhead{He}& 
\colhead{C}&
\colhead{N}&
\colhead{O}&
\colhead{Ne}&
\colhead{S}&
\colhead{Cl}&
\colhead{Ar}&
\colhead{Xe}&
\colhead{Ba}
}
\startdata
This work (RL)              &{\bf 11.04}    &{\bf 9.37} &$\cdots$ &{\bf 8.42} &$\cdots$ &$\cdots$&$\cdots$&$\cdots$&$\cdots$&$\cdots$\\
This work (CEL)             &$\cdots$ &{\bf 9.02} &{\bf 7.56} &{\bf
 8.18} &{\bf 6.43} &{\bf 5.13}&{\bf 3.88}&{\bf 4.56}&{\bf $>$2.75}&{\bf $<$2.51}\\
Kwitter et al. (2003) &11.08 &$\cdots$ &7.76 &8.30 &6.60 &5.30&$\cdots$&4.30&$\cdots$&$\cdots$\\
Henry et al. (1996)$^{a}$ &11.00 &8.68 &7.76 &8.40 &6.44 &$\cdots$&$\cdots$&$\cdots$&$\cdots$&$\cdots$\\
Torres-Peimbert \& Peimbert (1979, $t^{2}$=0)&   10.99  &9.39$^{b}$ &7.75&8.37&6.68&$\cdots$&$\cdots$&$\cdots$&$\cdots$&$\cdots$\\
Torres-Peimbert \& Peimbert (1979, $t^{2}$=0.035)&10.99 &9.39$^{b}$ &7.87&8.50&6.80&$\cdots$&$\cdots$&$\cdots$&$\cdots$&$\cdots$
\enddata
\tablecomments{The CEL abundances with $t^{2}\neq$ 0 by us are listed in Table \ref{t2}.}
\tablenotetext{a}{Derived from photo-ionization modeling.}
\tablenotetext{b}{Derived from C\,{\sc ii} $\lambda$4267 {\AA} line.}
\end{deluxetable*}

\begin{deluxetable}{@{}l@{\hspace{3pt}}c@{\hspace{5pt}}c@{\hspace{5pt}}c@{\hspace{5pt}}c@{\hspace{5pt}}c@{\hspace{5pt}}l@{}}
\tablecolumns{7}
\centering
\tablecaption{C, Ar, Xe, and Ba abundances in PNe.\label{xe}}
\tablewidth{\columnwidth}
\tablehead{
\colhead{Nebulae} &
\colhead{$[$C/Ar$]$}&
\colhead{$[$Xe/Ar$]$}&
\colhead{$[$Ba/Ar$]$}&
\colhead{$[$Ba/Xe$]$}&
\colhead{$[$Ar/H$]$}&
\colhead{Ref.}
}

\startdata
IC418 & +0.94 & +0.94 &+0.33&--0.61 &--0.54 &(1),(2)  \\ 
IC2501 & +0.66 & --0.03 &--0.06 &+0.09 &--0.27 &(1),(3),(4)  \\ 
IC4191 & +0.93 & +0.50 &+0.83 &+0.33 &--0.51 &(1),(5),(6)  \\ 
M3-15 & +0.23  & +1.15 &$\cdots$ &$\cdots$& --0.25 &(7),(8)  \\ 
NGC2440 & +0.71  & --0.36 &+0.79&+1.15& --0.24 &(1),(9)  \\ 
NGC5189 & --0.10  & --0.08 &$\cdots$ &$\cdots$& --0.06 &(7),(8)  \\ 
NGC7027 & +0.91  & +0.93 &+0.43&--0.50& --0.20 &(1),(10)  \\ 
PC14 & +0.43  & +0.13 &$\cdots$ &$\cdots$& --0.17 &(7),(8)  \\ 
Pe1-1 & +0.74  & +1.00 &$\cdots$ &$\cdots$&--0.23 &(7),(8)  \\ 
BoBn1 & +2.85  & $<$+1.95 &$>$+2.34 &$>$+0.38&--2.22 &(11)  \\ 
H4-1 & +2.62  & $>$+2.47 &$<$+2.32 &$<$--0.18&--1.99 &(7)  
\enddata
\tablecomments{The abundances are estimated from CELs, except
 for C in IC4191 which is estimated using recombination lines.
}
\tablerefs{
(1) \citet{2007ApJ...659.1265S};
(2) \citet{Pottasch:2004aa};
(3) \citet{2004AJ....127.2284H};
(4) \citet{Rola:1994aa};
(5) \citet{Pottasch:2005aa};
(6) \citet{2004MNRAS.353..953T};
(7) This work;
(8) \citet{Garcia-Rojas:2012aa};
(9) \citet{Bernard-Salas:2002aa};
(10) \citet{Zhang:2005aa};
(11) \citet{Otsuka:2010aa}
}
\end{deluxetable}

The elemental abundances were estimated using an ionization correction
factor, ICF(X), which is based on the ionization potential. The ICF(X) for
each element is listed in Table \ref{icf}.

The He abundance is the sum of the He$^{+}$ and He$^{2+}$ abundances. 
We assume that the C abundance is the sum of the
C$^{+}$, C$^{2+}$, C$^{3+}$, and C$^{4+}$ abundances, and we 
corrected for the CEL C$^{4+}$ and the RL C$^{+}$.
The N abundance is the sum of N$^{+}$, N$^{2+}$, and N$^{3+}$, and we
corrected for the N$^{2+}$ and N$^{3+}$ abundances. 
The O abundance is the sum of the O$^{+}$, O$^{2+}$, and O$^{3+}$ abundances.
For the RL O abundance, we estimated the O$^{+}$ and O$^{3+}$ abundances.
The Ne abundance is the sum of the Ne$^{+}$, Ne$^{2+}$, and Ne$^{3+}$ abundances,
and we corrected for the Ne$^{3+}$ abundance. The S abundance is the sum
of the S$^{+}$, S$^{2+}$, and S$^{3+}$ abundances. The Cl abundance is the sum of
the Cl$^{+}$, Cl$^{2+}$, Cl$^{3+}$, and Cl$^{4+}$ abundances,
correcting for Cl$^{+}$, Cl$^{3+}$, and Cl$^{3+}$. 
The Ar abundance is the sum of the Ar$^{+}$, Ar$^{2+}$, Ar$^{3+}$, 
and Ar$^{4+}$ abundances, correcting for the Ar$^{+}$ 
and Ar$^{4+}$ abundances. The Xe abundance is the sum of 
the Xe$^{+}$, Xe$^{2+}$, Xe$^{3+}$, and Xe$^{4+}$ abundances,
correcting for the Xe$^{+}$, Xe$^{3+}$, and Xe$^{4+}$ abundances. We did not correct for the 
ionization of Ba.

The resulting elemental abundances are listed in Table \ref{abund}. 
The types of emission lines used for the abundance estimations are 
specified in the second column. The number densities of each element 
relative to hydrogen are listed in the third column. The fourth 
column lists the number densities and the fifth column lists the number densities relative to the solar value. 
The last two columns are the solar abundances and the adopted ICF
values. We referred to \citet{2009ARA&A..47..481A} for N and Cl, and 
\citet{2003ApJ...591.1220L} for the other elements.

The [C/O] abundances are +1.25$\pm$0.19 dex from the RL and +1.14$\pm$0.20
dex from the CEL, therefore H4-1 is an extremely C-enhanced PN. 
However, as shown in the fifth column of Table \ref{abund}, 
the C and O abundances are slightly different between RLs and CELs. 
The C and O abundance discrepancies could be explained by small
temperature fluctuations in the nebula. We discuss the C and O abundance
discrepancies in the next section.

In Table \ref{past}, we compiled results for H4-1. We improved the C, O,
Ne, S, Ar abundances and newly added the Cl and Xe abundances, thanks 
to high-dispersion Subaru/HDS spectra and the detection of many different ionization 
stage ions. The large discrepancy in C abundance between the RL
\citep{Torres-Peimbert:1979aa} and the CEL \citep{Henry:1996aa} is reduced as a result of our work.

In Table \ref{xe}, we summarize the Xe abundances in 11 Galactic PNe. 
Using the line lists from \citet{Garcia-Rojas:2012aa}, we estimated the
Xe abundances of M3-15, NGC5189, PC14, and Pe1-1 in {\te}({\oiii}) and {\Ne}({\cliii}). 
For the above PNe, we removed the contribution of He\,{\sc ii}$\lambda$5846.7 to 
[Xe\,{\sc iii}]$\lambda$5846.8 using method I as described in Section 3.3.
The Xe abundance in H4-1 is the highest among these PNe. In the next
section, we also discuss how much Xe in H4-1 is a product of AGB nucleosynthesis in the progenitor.

\section{Discussion}
\subsection{C and O abundance discrepancies}

\begin{deluxetable}{@{}lclc@{}}
\tablecolumns{4}
\centering
\tablecaption{The Ionic and Elemental Abundances From CELs in $t^{2}$=0.03.\label{t2}}
\tablewidth{\columnwidth}
\tablehead{
\colhead{X$^{\rm m+}$} &
\colhead{(X$^{\rm m+}$/H$^{+}$)$_{t^{2} \neq 0}$}&
\colhead{X$^{\rm m+}$} &
\colhead{(X$^{\rm m+}$/H$^{+}$)$_{t^{2} \neq 0}$}
}
\startdata
C$^{+}$	      & 3.98(--4)$\pm$5.06(--5) &Ne$^{2+}$  &     2.42(--6)$\pm$1.91(--7)\\
C$^{2+}$&	6.25(--4)$\pm$9.38(--5) &S$^{+}$   &      2.27(--8)$\pm$1.37(--9)\\
C$^{3+}$&	8.96(--5)$\pm$1.85(--5) &S$^{2+}$   &     8.63(--8)$\pm$9.79(--9)\\
N$^{+}$	&	1.21(--5)$\pm$5.77(--7) &S$^{3+}$   &     3.48(--8)$\pm$4.09(--9)\\
O$^{+}$	&	4.83(--5)$\pm$3.78(--6) &Cl$^{2+}$  &     4.37(--9)$\pm$5.21(--10)\\
O$^{2+}$&	1.17(--4)$\pm$6.82(--6) &Ar$^{2+}$   &    2.05(--8)$\pm$1.12(--9)\\
O$^{3+}$   &     7.80(--6)$\pm$7.23(--7) &Ar$^{3+}$  &     9.25(--9)$\pm$9.58(--10)\\ 
Ne$^{+}$     &   4.66(--7)$\pm$6.81(--8)& Xe$^{2+}$  &     $>$3.61(--10)\\
\hline\hline
\noalign{\smallskip}
X &$\log$(X/H)$_{t^{2} \neq 0}$+12  &X &$\log$(X/H)$_{t^{2} \neq 0}$+12\\
\noalign{\smallskip}
\hline
C  &9.05$\pm$0.18  &S& 5.15$\pm$0.03\\
N  &7.64$\pm$0.04  &Cl& 3.95$\pm$0.13\\
O  &8.24$\pm$0.02  &Ar& 4.62$\pm$0.11\\
Ne &6.48$\pm$0.10  &Xe& $>$2.84
\enddata
\tablecomments{The RL elemental C and O abundances in $t^{2}$=0.03 are
 9.35$\pm$0.13 and 8.40$\pm$0.12, respectively (See text in detail).}
\end{deluxetable}

The RL to CEL abundance ratio, also known as the abundance discrepancy 
factor (ADF), in C$^{2+}$, C$^{3+}$, and O$^{2+}$ 
is 2.20$\pm$0.45, 4.36$\pm$0.81, and 1.75$\pm$0.36, respectively. 
For most PNe, ADFs are typically between 1.6 and 3.2 \citep[see][]{Liu:2006aa}.

The value of {\te}({\hei}) is comparable to the value of {\te}(BJ) within estimation error. 
According to \citet{2005MNRAS.358..457Z}, if {\te}({\hei}) is lower 
than {\te}(BJ), then the chemical abundances in the nebulae have two different 
abundance patterns. Otherwise, the abundance discrepancy between the CELs and the RLs is caused by temperature fluctuations within the
nebula.

We attempted to explain the discrepancies in C and O abundance by including 
temperature fluctuations in the nebula. \citet{Peimbert:1993aa} extended 
this effect to account for RL and CEL O$^{2+}$ abundance discrepancies by
introducing the mean electron temperature $T_{0}$ and 
the electron temperature fluctuation parameter $t^{2}$, as follows,
\begin{eqnarray}
&&
T_{0}({\rm X}^{i+}) = \frac{\int T_{\epsilon}n_{\epsilon}N({\rm X}^{i+})~dV}{\int n_{\epsilon}N({\rm X}^{i+})~dV},\\
&&
t^{2} = \frac{\int (T_{\epsilon} - T_{0})^{2}n_{\epsilon}N({\rm X}^{i+})~dV}
{T_{0}\int n_{\epsilon}N({\rm X}^{i+})~dV}.
\end{eqnarray}

Using this temperature fluctuation model, we attempted to explain the discrepancy in
C$^{2+,3+}$ and O$^{2+}$ abundance.

Following the methods of \citet{Peimbert:1971aa} and \citet{Peimbert:2003aa}, 
we estimated the mean electron temperatures and $t^{2}$. 
When $t^{2}$ $\ll$ 1, the observed {\te}({\oiii}), {\te}({\nii}), 
and {\te}(BJ) are written as follows,
\begin{eqnarray}
&&T_{\epsilon}({\rm [O\,{\sc III}]}) = T_{0,h}\left[1 +
     \frac{1}{2}\left(\frac{91\,300}
{T_{0,h}} - 3 \right)t_{h}^{2}\right], \label{tfo3}\\
&& \nonumber\\
&&T_{\epsilon}({\rm [N\,{\sc II}]}) = T_{0,l}\left[1 +
     \frac{1}{2}\left(\frac{69\,070}
{T_{0,l}} - 3 \right)t_{l}^{2}\right], \label{tfn2}\\
&&T_{\epsilon}({\rm BJ}) = T_{0}\left(1 - 1.67~t^{2}\right), \label{balje}
\end{eqnarray}
\noindent
where $T_{0,h}$ \& $t_{h}^{2}$ and $T_{0,h}$ \& $t_{l}^{2}$ 
are the average temperatures and temperature fluctuation parameters 
in high- and low-ionization zones, respectively; 
C$^{+}$, N$^{+}$, O$^{+}$, and S$^{+}$ are in low-ionization zones 
and the other elements are in high-ionization zones (See Table \ref{icf}). 

Based on the assumption that $t^{2}$=$t_{h}^{2}$=$t_{l}^{2}$, we found that
$T_{0,h}$=12\,590$\pm$340 K, $T_{0,l}$=10\,760$\pm$290 K, 
$T_{0}$=12\,590$\pm$1100 K, and $t^2$=0.030$\pm$0.007 
minimize the ADFs in C$^{2+,3+}$ and O$^{2+}$ using 
the combination of equations (\ref{tfo3})-(\ref{balje}). 
Our determined $t^{2}$ is in agreement with
\citet{Torres-Peimbert:1979aa}, who determined $t^{2}$=0.035.

Next, the average line-emitting temperatures for each line at 
$\lambda$, $T_{\epsilon}(\lambda)$, and $T_{\epsilon}$({\hb}) can be
written by following the method of \citet{Peimbert:2004aa}: 
\begin{eqnarray}
&&
T_{\epsilon}(\lambda)=T_{0,h,l} \nonumber\\ 
&&\times\left\{1+\left[\frac{(\Delta{E}/kT_{0,h,l})^2-3(\Delta{E}/kT_{0,h,l})+0.75}
{(\Delta{E}/kT_{0,h,l})-0.5}\right]\frac{t^2}{2}\right\},\\
&& \nonumber\\
&&T_{\epsilon}({\rm H\beta}) = T_{0,h,l}\left(1 -
					 \frac{1.633}{2}t^{2}\right),
\end{eqnarray}
\noindent where $T_{0,h,l}$ is 12\,590$\pm$340 K for the elements in high ionization
zone and is 10\,760$\pm$290 K for those in low ionization zone, 
the  $\Delta{E}$ is the difference energy between the upper and lower
level of of the target lines, $k$ is Boltzmann constant,
respectively. Finally, the ionic abundances can be estimated by
following the method of \citet{Peimbert:2004aa}, which accounts for the effects of fluctuations in temperature:
\begin{eqnarray}
&&
\frac{N({\rm X}^{m+})_{t^{2} \neq 0.00}}{N({\rm
X}^{m+})_{t^{2} = 0.00}} = 
\frac{T_{\epsilon}({\rm
H\beta})^{-0.87}T_{\epsilon}(\lambda)^{0.5}}{T_{\epsilon}([\rm O\,{\sc
III}],[N\,{\sc II}])^{-0.37}}
\nonumber \\
&& \nonumber\\
&&
\times \exp\left[\frac{\Delta E}{kT_{\epsilon}(\lambda)} -
	  \frac{\Delta E}{kT_{\epsilon}([\rm O\,{\sc III}],[N\,{\sc II}])}\right].
\end{eqnarray}

The resulting CEL ionic and elemental abundances are summarized in Table
\ref{t2}. The ADFs are slightly improved for C$^{2+}$ and O$^{2+}$,
with values of 1.94$\pm$0.45 and 1.45$\pm$0.30, respectively, whereas the ADF for
C$^{3+}$ remains high, with a value of 4.31$\pm$1.03. Because most of 
the ionic C abundances are in a doubly ionized stage, and 
the CEL C$^{2+}$ abundance approaches that of the RL C$^{2+}$, the
elemental CEL C abundance is very close to the RL C abundance
(9.35 dex for the RL and 9.05 dex for the CEL), resolving the C
abundance discrepancy. The elemental O
abundances agree within error, 8.40 for the RL and 8.24 for the CEL. 

Two difference abundance patterns could cause the ionic C$^{2+,3+}$ and 
O$^{2+}$ abundances. However, even if these abundance patterns appear in H4-1, the difference between
these ionic abundances would be negligible.

\subsection{Comparison with theoretical model}

\begin{deluxetable}{@{}l@{\hspace{5pt}}c@{\hspace{5pt}}c@{\hspace{5pt}}
c@{\hspace{5pt}}c@{\hspace{5pt}}c@{\hspace{5pt}}c@{\hspace{5pt}}c@{\hspace{5pt}}c@{}}
\tablecolumns{9}
\centering
\tablecaption{
Comparison of the observed abundances with the model prediction. \label{modelabun}}
\tablewidth{\columnwidth}
\tablehead{
\colhead{References}&
\colhead{C}&
\colhead{N}&
\colhead{O}&
\colhead{Ne}&
\colhead{S}&
\colhead{Ar}&
\colhead{Xe}&
\colhead{Ba}
}
\startdata
0.9 $M_{\odot}$                  &9.04 &7.57  &7.63 &7.87 &5.00 &4.28 &1.56   &2.37\\
initial abundances &6.28 &5.68  &6.54 &5.78 &4.97 &4.25 &0.01&0.00\\ 
\hline
2.0 $M_{\odot}$              &9.55 &6.84  &7.93 &8.66 &5.34 &4.57 &2.11   &2.52\\ 
   + $r$+2.0 dex\\
initial abundances &6.06 &4.64  &6.82 &6.06 &5.26 &4.49 &1.90&1.29\\ 
\hline
This work ($t^{2}$=0)&9.02 &7.56  &8.18 &6.43 &5.13 &4.56&$>$2.75&$<$2.51
\enddata
\end{deluxetable}

The large enhancement of the Xe abundance is the most remarkable finding for
H4-1. In this section, we compare the observed abundances with the
nucleosynthesis models for the initially 0.9 $M_{\odot}$ and 2.0
$M_{\odot}$ single stars with [X/H]=--2.19 from \citet{Lugaro:2012aa}, and 
we verify whether the observed Xe abundance could be explained using
their models. From the galaxy chemical evolution model of 
\citet{Kobayashi:2011aa}, an [Fe/H] ratio of about --2.30 in 
H4-1 is estimated using the relation [Fe/H]=[Ar/H]-0.30. Therefore, 
the results of \citet{Lugaro:2012aa} are good comparisons.

For the initially 0.9 $M_{\odot}$ star model, \citet{Lugaro:2012aa} adopted scaled-solar abundances as 
the initial composition for all elements from C to Pb. Both the initial
Xe and Ba abundances are $\sim$0, meaning that the final predicted abundances
of these two elements are pure products of the $s$-process in their
progenitors. In the 2.0 $M_{\odot}$ model, they initially adopted the
[$r$/Fe]=+2. In Table \ref{modelabun}, we compare our results with the
predictions by \citet{Lugaro:2012aa}. The initial abundances are also
listed. The models of \citet{Lugaro:2012aa} include a partial 
mixing zone (PMZ) of 2(--3) $M_{\odot}$ that produces a $^{13}$C pocket
during the interpulse period and releases extra free neutrons ($n$) through
$^{13}$C($\alpha$,{\it n})$^{16}$O to obtain $n$-process elements. 
The 0.9 $M_{\odot}$ and the 2.0 $M_{\odot}$ star models predict that 
the respective stars experienced 38 and 27 times thermal
pulses and the occurrence of a TDU.

The 0.9 $M_{\odot}$ model gives a good agreement with the observed C and
N abundances, however, there are discrepancies between the observed and the model predicted 
$\alpha$-elements. The production of Xe is mainly a result of the
$r$-process; \citet{Bisterzo:2011aa} reported that the solar Xe and Eu are
mainly from the $r$-process (84.6 $\%$ and 94 $\%$, respectively),
although the contribution from the $r$-process to the Xe is disputable.
The 0.9 $M_{\odot}$ model predicted the Xe/H abundance of 3.63(--11),
while the observed abundance in H4-1 is 5.62(--10). Therefore, most the xenon
in H4-1 is produced by the $r$-process in primordial SNe.

If the progenitor of H4-1 was formed in an $r$-process 
rich environment, the observed $\alpha$ elements and Xe could be 
explained. Compared to the 0.9 $M_{\odot}$ model, except for N and Ne, the $r$-process 
enhanced model for 2.0 $M_{\odot}$ stars explains the observed O, S, Ar, and Xe abundances well.
About 0.2-0.3 dex of the observed $\alpha$-elements are SN products.

Through comparison with the theoretical models, we suppose 
that the observed Xe is mostly synthesized by $r$-process in SNe. 
The observed abundances could be explained by the model 
for 2.0 $M_{\odot}$ stars with the initial [$r$/Fe]=+2.0.

\subsection{Comparison with CEMP stars and the evolution and evolution
  of H4-1}

\begin{figure}
\centering
\includegraphics[width=\columnwidth]{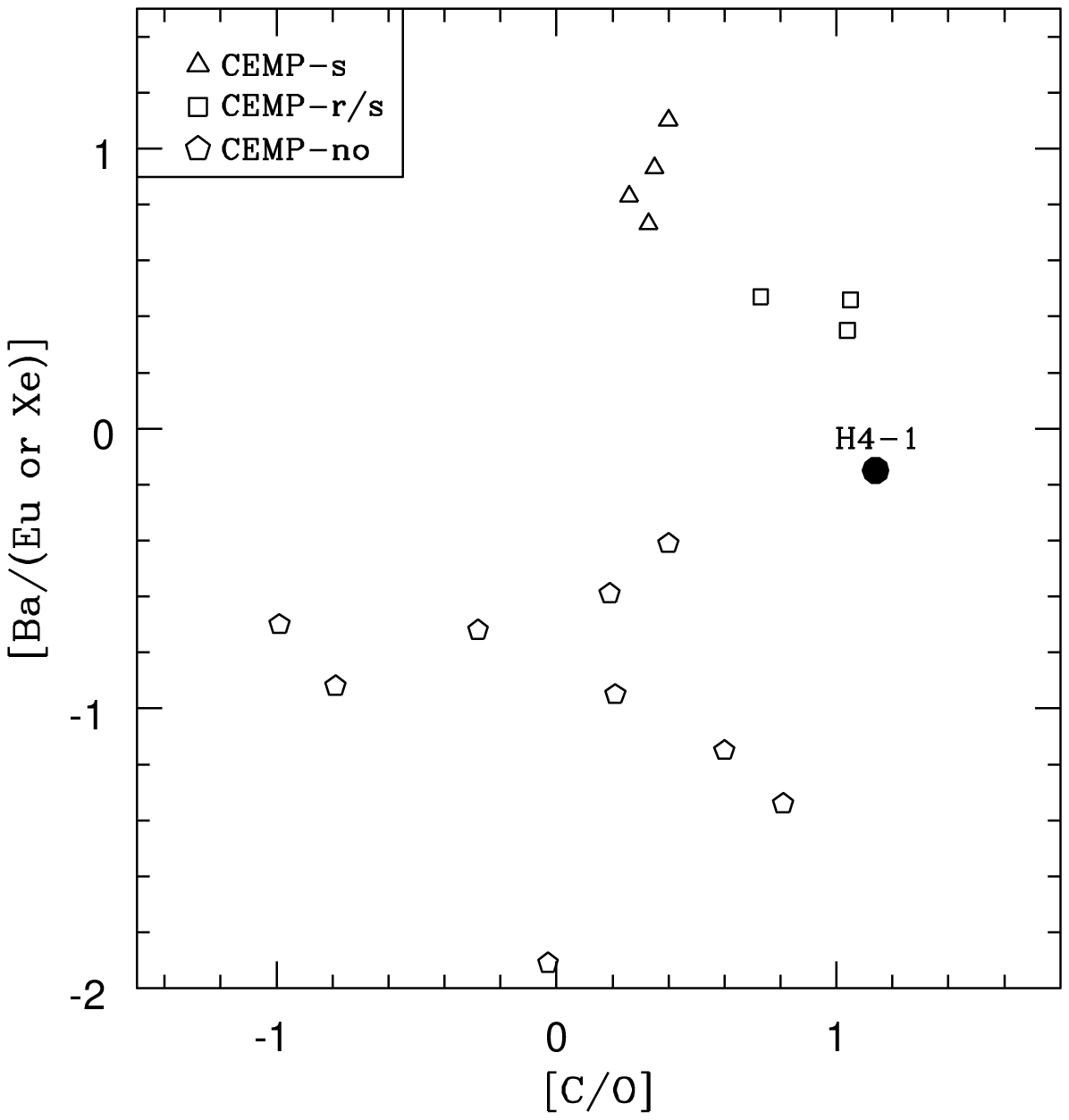}
\caption{The [Ba/(Eu or Xe)] versus [C/O] diagram. For H4-1, 
we used the [Ba/Xe] abundance. The [Ba/Xe] abundance of H4-1 is the
 upper limit. We used the abundances of CEMP stars
 provided by \citet{Suda:2011aa}. CEMP stars are classified as the criteria
 summarized in Table \ref{cemp}. The [Fe/H] of CEMP stars 
plotted in this diagram is from --4.75 to -2.03 in CEMP-$no$, from
--2.7 to --2.2 in CEMP-$s$, and from --2.72 to --2.03 in CEMP-$r$/$s$, respectively.
\label{baeuco}}
\end{figure}

\begin{deluxetable}{@{}ll@{}}
\tablecolumns{2}
\centering
\tablecaption{Classification of CEMP stars in this work. \label{cemp}}
\tablewidth{\columnwidth}
\tablehead{
\colhead{Class}&
\colhead{Criteria}
}
\startdata
CEMP-$s$     &[Fe/H]$\leq$--2,[C/Fe]$>$+1.0,[Ba/Fe]$>$+1.0,[Ba/Eu]$>$+0.5\\
CEMP-$r$/$s$ &[Fe/H]$\leq$--2,[C/Fe]$>$+1.0,[Ba/Fe]$>$+1,0$<$[Ba/Eu]$<$+0.5\\
CEMP-$no$     &[Fe/H]$\leq$--2,[C/Fe]$>$+1.0,[Ba/Fe]$<$0.0
\enddata
\tablecomments{In H4-1, the expected [Fe/H] and [Ba/Fe] are --2.3 and $<$+2.6 and the observed [C/Fe] and [Ba/Xe] are +2.83 and $<$--0.15,
respectively.}
\end{deluxetable}

Although the models by \citet{Lugaro:2012aa} explain the observed
abundances, it is difficult to explain the
evolutional time scale of H4-1 with a 2.0 $M_{\odot}$  
single-star evolution models, because such mass stars cannot survive in
the Galactic halo up to now.

Therefore, it is possible that H4-1 evolved from a binary, similar to
the evolution of CEMP stars, and its progenitor was 
polluted by SNe. Our definition of CEMP stars is summarized in Table
\ref{cemp}, following \citet{Beers:2005aa}. CEMP-$s$ stars are $s$-process rich, 
CEMP-$r$/$s$ are both $s$- and $r$-process rich. The CEMP stars which
are neither CEMP-$s$ nor CEMP-$r$/$s$ are classified as the CEMP-$no$. 
Since the expected [Fe/H] and [Ba/Fe] are --2.3 and $<$+2.6 and 
the observed [C/Fe] and [Ba/Xe] are +2.83 and $<$--0.15,
respectively, H4-1 can be classified as CEMP-$r$/$s$ or CEMP-$no$.

In Fig. \ref{baeuco}, we present the [Ba/(Eu or Xe)]
versus [C/O] diagram between H4-1 and CEMP stars. The data of CEMP
stars are taken from \citet{Suda:2011aa} and the classification is based on
Table \ref{cemp}. The metallicities of CEMP-$s$ and CEMP-$r$/$s$ stars 
are widely spread, while those of CEMP-$no$ are mostly 
$\lesssim$--3 \citep[e.g.,][]{2007ApJ...655..492A}. Fig. \ref{baeuco} 
indicates that H4-1 is similar to CEMP-$r$/$s$ and CEMP-$no$, 
assuming that the [Eu/H] abundance in H4-1 is similar to its [Xe/H] abundance.

Although H4-1 has a similar origin and evolution as CEMP-$r/s$
and CEMP-$no$ stars, the origin of these CEMP stars is a topic of much debate 
\citep{Lugaro:2012aa,Bisterzo:2011aa,2006A&A...451..651J,Ito:2013aa,Zijlstra:2004aa}. 
As proposed by \citet{Zijlstra:2004aa} to 
explain the abundances of $r/s$-rich stars, the progenitor of H4-1 
might be a binary composing of e.g, $\sim$0.8-0.9 $M_{\odot}$ and 
$\sim$5 $M_{\odot}$ (from their Fig. 2), which might evolve into a SN. 
However, there are problems on the N and Xe productions in such massive 
primary scenario for explanation of H4-1's evolution. For instance, the 
5 $M_{\odot}$ with $r$+0.4 dex models by \citet{Lugaro:2012aa} 
predicted highly enhanced N (9.11) by the hot bottom burning and low Xe abundances (0.99
dex). While, \citet{Suda:2013aa} argued that the inclusion of mass-loss
suppression in metal-poor AGB stars can inhibit such N-enhanced 
metal-poor stars. Therefore, at the present, we think that the primary
star should have never experienced the hot bottom burning.

If bipolar nebulae are created by stable mass-transfer during Roche lobe
overflow, the initial mass ratio of the primary to secondary is $\sim$1-2, 
according to \citet{Phillips:2000aa}. If this is the case for the bipolar nebula
formation in H4-1 and the secondary is $\sim$0.8-0.9 $M_{\odot}$, 
the primary star would be $\sim$0.8-1.8 $M_{\odot}$. For example, 
\citet{Otsuka:2010aa} explain the observed chemical abundances of BoBn1 
using the binary model composed of 0.75 $M_{\odot}$ + 1.5 $M_{\odot}$ stars.

\section{Summary}
We analyzed the multi-wavelength spectra of the halo PN H4-1 
from Subaru/HDS, \emph{GALEX}, \emph{SDSS}, and \emph{Spitzer}/IRS in 
order to to determine chemical abundances, in particular, $n$-capture 
elements, solve the C abundance discrepancy problem, and obtain insights
on the origin and evolution of H4-1. We determined the abundances of 10
elements based on the over 160 lines detected in those data. 
The C and O abundances were derived from both CELs and RLs. We found 
the discrepancies between the CEL and the RL abundances of C and O,
respectively and they can be explained by considering temperature 
fluctuation effect. The large discrepancy in the C abundance between 
CEL and RL in H4-1 was solved by our study. In HDS spectrum, we
detected the [Xe~{\sc iii}]$\lambda$5846 {\AA} line in H4-1 for the
first time. H4-1 is the most Xe enhanced PN among the Xe detected PNe. 
The observed abundances can be explained by a $\sim$2.0 $M_{\odot}$ 
single star model with initially [$r$/Fe]=+2.0 
of \citet{Lugaro:2012aa}. The observed Xe abundance would be
 a product of the $r$-process in primordial SNe. About 0.2-0.3 dex of
 the $\alpha$ elements are also the products by these SNe. The [C/O]-[Ba/(Eu or
 Xe)] diagram suggests that the progenitor of H4-1 shares the evolution
 with CEMP-$r$/$s$ and CEMP-$no$ stars. The progenitor of H4-1
 is a presumably binary formed in an $r$-process rich environment.

\section*{Acknowledgments}
We are grateful to the anonymous referee for a careful reading and valuable suggestions. 
We sincerely thank Takuma Suda and Siek Hyung for fruitful discussions on low-mass
AGB nucleosynthesis and PN abundances. This work is mainly based on data 
collected at the Subaru Telescope, which is operated by 
the National Astronomical Observatory of Japan (NAOJ).
This work is in part based on {\it GALEX} archive data
downloaded from the {\it MAST}. 
This work is in part based on archival data obtained 
with the {\it Spitzer Space Telescope}, which is operated by the 
Jet Propulsion Laboratory, California Institute of 
Technology under a contract with NASA. Support for this 
work was provided by an award issued by JPL/Caltech. 
Funding for the SDSS has been provided by the Alfred P. Sloan 
Foundation, the Participating Institutions, the National Aeronautics 
and Space Administration, the National Science Foundation, the U.S. 
Department of Energy, the Japanese Monbukagakusho, the Max Planck 
Society, and the Higher Education Funding Council for England.

\end{document}